\documentclass[aps,prd,showpacs,11pt]{revtex4}
\usepackage{graphicx}
\begin{document}
\def\la{{\langle}}
\def\ra{{\rangle}}
\def\vep{{\varepsilon}}
\newcommand{\beq}{\begin{equation}}
\newcommand{\eeq}{\end{equation}}
\newcommand{\beqa}{\begin{eqnarray}}
\newcommand{\eeqa}{\end{eqnarray}}
\newcommand{\da}{^\dagger}
\newcommand{\wh}{\widehat}

\title{Escape of photons from two fixed extreme Reissner-Nordstr\"om black holes}
\date{\today}

\author{Daniel Alonso$^{1,2}$,Antonia Ruiz$^{1,2}$ and Manuel S\'anchez-Hern\'andez$^{2}$}

\affiliation{
$^1$ Instituto Universitario de Estudios Avanzados (IUdEA) \\ en F\'\i sica At\'omica, Molecular y Fot\'onica \\
Universidad de La Laguna, Facultad de F\'\i sicas, La Laguna 38203, Tenerife, Spain \\
$^2$ Departamento de F\'{\i}sica Fundamental y Experimental, Electr\'onica y Sistemas. \\
Universidad de La Laguna, La Laguna 38203, Tenerife, Spain}


\begin{abstract}
We study the scattering of light (null geodesics) by two fixed extreme Reissner-Nordstr\"om black holes, in which
the gravitational attraction of their masses is exactly balanced with the electrostatic repulsion of their
charges, allowing a static spacetime. We identify the set of unstable periodic orbits that form part of the
fractal repeller that completely describes the chaotic escape dynamics of photons. In the framework of periodic
orbit theory, the analysis of the linear stability of the unstable periodic orbits is used to obtain the main
quantities of chaos that characterize the escape dynamics of the photons scattered by the black holes. In
particular, the escape rate which is compared with the result obtained from numerical simulations that consider
statistical ensembles of photons. We also analyze the dynamics of photons in the proximity of a perturbed black hole
and give an analytic estimate of the escape rate in this system.
\end{abstract}


\maketitle

\section{Introduction}

In general relativity, the nonexistence of an absolute time parameter introduces new aspects in the characterization 
of chaos with respect to the well-known Newtonian dynamics \cite{book}. 
Some relativistic systems in which the existence of chaos has been reported include charged particles in a magnetic 
field interacting with gravitational waves, spinning particles orbiting rotating and nonrotating black holes,
gravitational waves from spinning compact binaries, as well as particles in Majumdar-Papapetrou geometries.

Most of the studies of chaos around black holes have focused on the analysis of the qualitative changes in the 
dynamics of an isolated black-hole spacetime caused by a small perturbation due to external mass distributions
\cite{chandrasekhar1989,contopoulos1990a,dettmann1994,cornish1997,cornish2003a}, gravitational waves
\cite{bombelli1992}, spin-orbit and spin-spin coupling \cite{suzuki1997}, or magnetic fields \cite{karas1992}.
Recently, the projects to startup operative ground-base gravitational wave detectors (LIGO, VIRGO, GEO600, TAMA300, AIGO) and 
a planned laser interferometer space antenna (LISA) \cite{observ}, which will be able to detect the signals from gravitational 
wave sources, such as inspiralling compact binary systems of neutron stars or black holes, have increased the interest in 
the presence of chaos in the dynamics of binary black holes and its effects on the outgoing gravitational radiation
\cite{cornish2003}.

In this work we use the Majumdar-Papapetrou metric \cite{MPmetric} to analyze the dynamics of photons in the gravitational 
field of two extreme Reissner-Nordstr\"om black holes that are fixed in space due to the balance of their gravitational 
attraction and electrostatic repulsion. Although it is unlikely that this metric describes any astrophysical system, since 
in real universe black holes tend to rotate around their center of mass producing gravitational waves and do not possess  
overall electric charge, the chaotic scattering of photons in the Majumdar-Papapetrou static spacetime of nonrotating 
black holes with extreme electric charge still provides an interesting formal model that can be used to illustrate many 
of the features expected in more realistic systems.

In \cite{chandrasekhar1989} Chandrasekhar studied the scattering of radiation by two extreme Reissner-Nordstr\"om 
black-holes that are at finite distance apart. In contrast to the two center problem in Newtonian gravitation, 
in general relativity this two center gravitational problem is generally not integrable. An appendix in 
\cite{chandrasekhar1989} displays a set of null geodesics (photon's trajectories) in the meridian plane 
of the system and concludes that the dynamics is probably not separable.

Shortly after, Contopoulos \cite{contopoulos1990a} systematically studied the set of photon trajectories with zero 
angular momentum component along the axis that goes through the black-holes. In that case the motion is 
confined to a plane (the meridian plane). The study concluded that there are three types of non 
periodic motions. There are orbits that fall into one of the black-holes, with mass $M_1$ or mass $M_2$, these are
orbits of type $I$ and $II$, and there are orbits that escape to infinity, these are orbits of type $III$. 
The orbits of different types are separated by orbits that tend asymptotically to three 
main types of unstable periodic orbits. One kind of periodic orbits goes around one of the black-holes, either $M_1$ or 
$M_2$. A third type of unstable periodic orbits goes around the two black-holes.
Between two non periodic orbits of two different types there are orbits of a third type. 
In a related work Contopoulos {\sl et at.} \cite{contopoulos1993} analyze in detail the different types of periodic
orbits in the two fixed center system and compare families of these orbits in the Newtonian and the relativistic
problems. In this system chaos appears explicitly as the initial conditions of these types of orbits form a 
Cantor set. Further numerical evidence indicated that all the photon periodic orbits are unstable. 
In the same direction it was investigated the phase space trajectories in a multi-black-hole spacetime 
\cite{dettmann1994}, it was found that the chaotic geodesics are well described by Lyapunov exponents. 
All these works strongly indicate that the scattering of photons by two Reissner-Nordstr\"om black-holes held fixed 
is chaotic and that the set of periodic orbits is unstable.
The geometric analysis of the flow also showed the chaotic behavior of the relativistic null-geodesic motion in the
two black hole spacetime \cite{yurtsever}.
More recently Contopoulos {\sl et al.} have studied in detail the asymptotic curves from the periodic orbits, their
homoclinic and heteroclinic intersections and the basins of attraction of two black holes \cite{contopoulos2004}.

Our aim is to give a full description of the chaotic escape dynamics of photons in the two black-hole system in terms 
of the different physical indicators of chaos. We shall analyze the linear stability of the dynamics, in particular 
the unstable periodic orbits, and from that information we evaluate the escape rate associated with the equations of 
motion using the well known trace formula for hyperbolic flows of Cvitanovic and Eckhardt \cite{cvitanovic1991}. 
It turns out that the escape rate is given by the leading eigenvalue of the spectrum of the evolution operator 
\cite{gaspard1989}. In addition, we shall compare our results with numerical simulations that consider the time 
evolution of statistical ensembles of photons.

In Section II we describe the Chandrasekhas's model to analyze the photon dynamics under the gravitational field 
of two extreme Reissner-Nordstr\"om black holes.  In Section III we discuss the linear stability of the dynamics and 
the general methods that we will consider to obtain the escape dynamics of photons. 
Section IV introduces the formalism to study the time evolution of statistical ensembles of photons and their escape 
dynamics. In Section V we present the set of periodic orbits of the system in the meridian plane $(L_z=0)$ and give 
their periods and stretching factors. In this section we also illustrate the fractal repeller associated with the 
dynamics of photons in the two black-hole field. In Section VI we calculate the main quantities of chaos derived from 
the analysis of the linear stability of the unstable periodic orbits. The escape rate obtained from the periodic orbit 
theory is compared with the value obtained from numerical simulations that consider statistical ensembles of photons. 
In Section VII we analyze the dynamics in the proximity of a perturbed black hole and give an analytic estimate of 
the escape rate in this system. In Section VIII the main conclusions are put together. 

\section{Dynamical equations}

The model studied by Chandrasekhar in \cite{chandrasekhar1989} derives from the solutions of the Einstein-Maxwell equations 
to describe a problem analogous to the Newtonian arrangement of charged mass-points in which the mutual Coulomb repulsions 
are balanced with the gravitational attraction. These solutions are known as the Majumdar-Papapetrou solutions \cite{MPmetric}. 
They are obtained from an static solution of the Einstein-Maxwell equations \cite{chandrasekhar1983}. The metric of the 
Majumdar-Papapetrou solution of the Einstein-Maxwell equations is given by
\begin{equation}
\label{metric}
ds^2= dt^2/U^2
- U^2 (dx^2+dy^2+dz^2),
\end{equation}
where $(x,y,z)$ are the spatial coordinates. The function $U$, which depends only on the spatial coordinates, is a solution 
of the three-dimensional Laplace's equation
\begin{equation}
\nabla^2 U= \left(\frac{\partial^2}{\partial x^2}+\frac{\partial^2}{\partial y^2}+\frac{\partial^2}{\partial z^2} \right)U=0.
\end{equation}
Hartle and Hawking \cite{hartle1972} showed that for a function $U$ of the form
\begin{equation}
U=1+\sum_{i=1}^N \frac{M_i}{r_i},
\end{equation}
being $r_i=\sqrt{(x-x_i)^2+(y-y_i)^2+(z-z_i)^2}$, the Majumdar-Papapetrou solution corresponds to a system of $N$ extreme
Reissner-Nordstr\"om black holes with horizons at $(x_i,y_i,z_i)$ and with masses equal to their charges, $M_i=Q_i>0$.
The metric is everywhere regular except at the black-hole locations where there are coordinate singularities, 
as argued in \cite{chandrasekhar1989,chandrasekhar1983}.

We will consider a configuration of two black holes located at $(0, 0, \pm z_{bh})$. Using geometrized units 
(the speed of light in vacuum $c=1$ and the gravitational constant $G=1$) the function $U$ takes the form
\begin{equation}
\label{funcU}
U=1+\frac{M_1}{(x^2+y^2+(z-z_{bh})^2)^{1/2}}+\frac{M_2}{(x^2+y^2+(z+z_{bh})^2)^{1/2}}.
\end{equation}

The Lagrangian ${\cal L}$ associated with the metric (\ref{metric}) is defined by
\begin{equation}\label{lagran}
{\cal L}= \frac{\dot t ^2}{2U^2}-\frac{U^2}{2}(\dot x^2+\dot y ^2+\dot z^2),
\end{equation}
where the dot denotes the derivative with respect to the affine parameter $\mu$. The Hamiltonian ${\cal H}$ can be expressed
as
\begin{equation}
{\cal H}=\frac{1}{2} U^2 p_t^2-\frac{1}{2} U^{-2}(p_x^2+p_y^2+p_z^2),
\end{equation}
being the Hamilton's equations that dictate the geodesic motion associated with the metric (\ref{metric}):
\begin{eqnarray}
\label{eqnmotion}
&&\dot \alpha=\frac{d\alpha}{d\mu}=-U^{-2}p_\alpha \\
\nonumber
&&\dot p_\alpha=\frac{dp_\alpha}{d\mu}=-\frac{1}{2}\partial_\alpha(U^2-U^{-2}P^2)
\hspace*{1.5cm}(\alpha=x,y,z)
\nonumber.
\end{eqnarray}

According to the Hamilton's equations $p_t={\cal E}=constant$ and angular momentum along the $z$-axis 
$L_z=x p_y-y p_x=constant$. In the case of photons it follows also that 
${\cal H}=U^2 {\cal E}^2/2-U^{-2}(p_x^2+p_y^2+p_z^2)/2=0$ \cite{chandrasekhar1989,contopoulos1990a}, and therefore 
$p_x^2+p_y^2+p_z^2=P^2=U^4 {\cal E}^2$. Hence the problem scales with ${\cal E}$ and without loss of generality we can 
consider ${\cal E}=1$. 

The equations of motion with respect to the killing time $t$ can be written as:
\begin{eqnarray}
\label{eqnmotion2}
&&\alpha'=\frac{d\alpha}{dt}=\dot \alpha\frac{d\mu}{dt}=-U^{-4}p_\alpha \\
\nonumber
&&p_\alpha'=\frac{dp_\alpha}{dt}=\dot p_\alpha\frac{d\mu}{dt}=-2U^{-1}\partial_\alpha U
\hspace*{1.5cm}(\alpha=x,y,z)
\nonumber.
\end{eqnarray}
The sets of equations (\ref{eqnmotion}) and (\ref{eqnmotion2}) are related by the transformation 
\begin{equation}
\frac{d\,}{d\mu}=U^2\frac{d\,}{dt}
\end{equation}
between the affine parameter $\mu$ and the killing time $t$. 
Although the solutions of these two systems induce different flows, $\Phi^{\mu}$ and $\Phi^{t}$ respectively, and therefore 
lead to different values in some properties that depend on the flow, the spatial distribution of their corresponding 
set of orbits do not change, and neither are modified the critical elements of the flow, in particular the periodic orbits.

Due to the conservation of $L_z$, the motion for $L_z=0$ is restricted to a plane (the meridian plane). We shall take this 
simplification and study the dynamics in the meridian plane.

\section{Linear stability}

We are particularly interested in the stability analysis of the solutions of the equations of motion, i.e. how 
is they behave after being slightly perturbed. To fix the notation and make the paper self contained, in this section 
we introduce the basic concepts of linear stability. First we shall discuss this problem in a general context. Then the 
application to the two black hole system is straightforward. 

Let us consider a Hamiltonian system with $f$ degrees of freedom. The $2f$-dimensional phase-space can be denoted by 
$X=(X_1,X_2,\cdots,X_{2f})$, being $(X_1,\cdots,X_{f})$ the generalized coordinates and $(X_{f+1},\cdots,X_{2f})$ 
the conjugated momenta. The Hamilton's equations for a Hamiltonian ${\cal H}$ can be expressed as
\begin{equation}
\dot X= \Sigma \cdot\frac{\partial {\cal H}}{\partial X}\,\,,
\end{equation}
being $\dot X=dX/d\tau$, with $\tau$ the affine parameter $\mu$ or the killing time $t$, and $\Sigma$ a $2f \times 2f$
antisymmetric matrix of the symplectic form
\begin{equation}
\Sigma= \left( \begin{array}{cc} {\sl 0} & {\sl 1} \\ -{\sl 1} & {\sl 0}
\end{array} \right),
\end{equation}
with ${\sl 0}$ and ${\sl 1}$ the $f \times f$ null and unit matrices respectively.

A solution $X_\tau$ of these differential equations is a curve in phase-space corresponding to a given initial 
condition $X_0$. The stability analysis of this solution involves the study of the time evolution of a small 
perturbation $\delta X$ with respect to $X_\tau$. Considering $Y_\tau=X_\tau+\delta X$, it follows that 
$\delta X$ satisfies the initial value problem
\begin{equation}\label{eqL}
\delta \dot X= {\sf L}(\tau) \cdot \delta X,
\end{equation}
with ${\sf L}(\tau)= \Sigma \cdot \frac{\partial^2 {\cal H}}{\partial X^2}|_{X_\tau}$. This is a linear initial 
value problem with time-dependent coefficients. For a given initial perturbation $\delta X_0$ it has the solution
\begin{equation}
\label{pereq}
\delta X_\tau={\sf M}(X_0,\tau) \cdot\delta X_0.
\end{equation}
The $2f\times 2f$ matrix ${\sf M}(X_0,\tau)$ is usually referred as the {\it fundamental matrix} and gives the time evolution 
of an initial displacement $\delta X_0$. The fundamental matrix satisfies the differential equation
\begin{equation}\label{eqM}
\dot {\sf M}(X_0,\tau)= {\sf L}(\tau) \cdot {\sf M}(X_0,\tau)\,, \,\,\hbox{with} \,\,{\sf M}(X_0,0)={\sf 1}.
\end{equation}

The behavior of the perturbation $\delta X$ can be analyzed in terms of the Lyapunov exponents of $X_\tau$, which 
measure the rate of exponential separation or approach of initially infinitely close trajectories. The Lyapunov exponent
associated with the unit vector ${\bf e}_j=\delta X_j/|\delta X_j|$ along the direction of $\delta X_j$ is given by 
\begin{equation}
\lambda(X_0,{\bf e}_j)=\lim_{\tau \to \infty} \frac{1}{\tau}\ln | {\sf M}(X_0,\tau)\cdot{\bf e}_j |.
\end{equation}
The number of Lyapunov exponents is equal to the phase-space dimension. Positive Lyapunov exponents indicate 
dynamical instability of trajectories in phase space, and therefore an extreme sensitivity to the initial conditions.

For a periodic orbit $X_\tau$, the Lyapunov exponents are directly related to the stability of this solution. 
The Lyapunov exponents can be degenerate with multiplicity $m_i$, in the sense that several of them have the same value. 
They are ordered as $\lambda_1 \ge\lambda_2\ge \cdots \lambda_{\nu}$ with  $\sum\limits_{i=1}^{\nu} m_i=2f$. The fundamental 
matrix has a symplectic structure and therefore satisfies the relation ${\sf M}^{T} \Sigma {\sf M}=\Sigma$ \cite{arnold1968}. 
Due to this property all the Lyapunov exponents in Hamiltonian systems are grouped in pairs of equal absolute value and
opposite sign $\{\lambda_i,-\lambda_i\}_{i=1,\cdots,f}$. Furthermore, 
perturbations along the direction of the flow give a zero Lyapunov exponent, and due to the pairing property another one is 
also zero, which is a consequence of the conservation of energy \cite{gaspard1998,gaspard1995}. Thus at least two exponents 
are zero.

The matrix ${\sf M}(X_0,\tau)$ corresponding to a periodic orbit is also periodic with the same period. 
In general $X_\tau=X_{\tau+T}$, with $T=rT_p$ (r=1,2,$\cdots$) and $T_p$ the primitive period of the orbit. 
The linear stability of a periodic orbit is determined by the eigenvalues of ${\sf M}(X_0,\tau)$ over one 
primitive period, i.e. ${\sf M}(X_0,T_p)$. These eigenvalues, called {\it stretching factors} or {\it stability eigenvalues},
are obtained from the secular determinant
\begin{equation}
{\hbox {det}} \left[ {\sf M}(X_0,T_p)-\Lambda {\sf 1} \right]=0.
\end{equation}

Due to the symplectic property of the fundamental matrix the eigenvalues $\Lambda_j$ are also grouped in pairs. 
From the previous discussion two of them are equal to one. One of 
them corresponds to perturbations along the periodic orbit and the other to perturbations that are perpendicular to the energy 
shell. The Lyapunov exponents associated with a periodic orbit are given in terms of the stretching factors as 
$\lambda_i=\ln |\Lambda_i|/T_p$.

The stability of a particular motion is determined by the location of the stretching factor in the complex plane:
$|\Lambda_i| >1$ is a signature of instability and corresponds to the unstable manifold of the periodic orbit, 
$|\Lambda _i|<1$ is related to the stable manifold and $|\Lambda _i|=1$ corresponds to the center manifold of the orbit.

A straightforward application, which is directly related to our problem, is the study of the dynamics of two-dimensional 
systems. In these systems a periodic orbit is called {\it hyperbolic} when $\Lambda > 1 >\Lambda^{-1} > 0$, 
{\it hyperbolic with reflection} when $\Lambda <-1 <\Lambda^{-1}<0$, and  {\it elliptic} when $|\Lambda|=1$ \cite{arnold1968}.
The orbit stability changes whenever $\Lambda$ crosses the unit circle, and generally a bifurcation may occur in which
some periodic orbits can disappear or emerge. In autonomous systems the energy is the main bifurcation parameter. However, 
in the two black-hole system no bifurcation is expected as the photon dynamics scales with the energy.

A periodic orbit is characterized by its primitive period and its stretching factors $\Lambda_i$. We shall see below how 
this information can be used to analyze in detail the escape dynamics of photons scattered by two black-holes. 
In general the Lyapunov exponents and the periods of the periodic orbits are not invariant under time-transformations, 
but the stretching factor are invariant, as noted in \cite{dettmann1994,cornish2003a,cornish2001,cornishgibbons}. 
In particular Motter \cite{motter2003a} showed that if we consider a transformation of the form $d\mu=f(X)dt$, the Lyapunov 
exponents transform according to $\lambda_i^{\mu}=\lambda_i^t/<f>_t$, where 
$<f>_t=lim_{t \rightarrow\infty}(1/t)\int_0^t f(X(t')) dt'$. For a given periodic orbit $p$ the Lyapunov exponents over 
a period $T_{\mu}$ in the $\mu$-time and over a period $T_t$ in the $t$-time are related by 
$\lambda_i^{T_{\mu}}=\lambda_i^{T_t}/<f>^*_{T_t}$ with $<f>^*_{T_t}=(1/T_t)\int_0^{T_t} f(X_p(t')) dt'=T_{\mu}/T_t$. 
From these relations it can be concluded that $\lambda_i^{T_{\mu}}T_{\mu} =\lambda_i^{T_t} T_t$ which implies 
the invariance of the stretching factors, $\Lambda_i^{T_{\mu}}=\Lambda_i^{T_t}$.

\section{Evolution of the density of photons and their decay dynamics}

The motion of a single photon in the meridian plane is chaotic and it is useful, in order to characterize its escape 
dynamics from the black holes, to consider statistical ensembles that provide a probabilistic description of the process. 
Therefore, instead of analyzing a single photon orbit, we consider an ensemble described by some density $\rho$. 
For a given Hamiltonian ${\cal H}$ the time evolution of a density $\rho$ is dictated by the Liouville equation
\begin{equation}
\label{liouville}
\partial_\tau \rho= \hat L \rho,
\end{equation}
with the Liouville operator $\hat L$ given by the Poisson bracket as $\hat L=\{{\cal{H}},\rho\}$. Assuming a 
time-independent Hamiltonian, the solution of this linear equation can be expressed as
\begin{equation}
\label{liouville1}
\rho_\tau=e^{\hat L \tau} \rho_0,
\end{equation}
being $\rho_0$ the initial density. From (\ref{liouville1}), the average over the ensemble of any observable $A$ at time 
$\tau$ can be computed as;
\begin{equation}
\left< A \right>_\tau=\int dX \, A(\Phi^\tau X) \rho_0(X),
\end{equation}
being $\Phi^\tau$ the flow that maps an initial condition $X_0$ into $X_\tau$, according to
$X_\tau=\Phi^\tau X_0$. This last expression can also be written in the form
\begin{equation}\label{frobenius}
\left< A \right>_\tau=\int dX dY\, A(X)\delta (X-\Phi^\tau Y)\rho_0(Y),
\end{equation}
where the Dirac distribution $\delta (X-\Phi^\tau Y)$ represents the conditional probability density of a 
trajectory being at position $X$ at time $\tau$ given the initial position $Y$. Indeed the Dirac distribution defines 
the kernel of the evolution operator. The equation (\ref{frobenius}) defines two self adjoint operators as
\begin{equation}
\left< A \right>_\tau= 
\left< \hat P^{\dagger \tau} A | \rho_0 \right>=
\left< A| \hat P^{\tau}\rho_0 \right>.
\end{equation}
The evolution of a probability density is ruled by the {\it Frobenius-Perron} equation \cite{gaspard1998,gaspard1995}
\begin{equation}
\rho_\tau(X)=\hat P^\tau \rho_0(X) \equiv \int dY \,\delta (X-\Phi^\tau Y) \rho_0(Y) 
\end{equation}
and the time evolution of the observable is dictated by the Koopman operator \cite{koopman1931} defined by
\begin{equation}
A_\tau(Y) = \hat P^{\dagger \tau } A(Y) \equiv \int dX \, \delta (X-\Phi^\tau Y) A(X).
\end{equation}

In the case of an invertible and conservative flow $\Phi^\tau$ the Frobenius-Perron operator reduces to
$\rho_\tau(X)=\hat P ^\tau \rho_0(X) = \rho_0(\Phi^{-\tau} X)$. In open systems the trajectories that are initially 
confined in a bounded region in phase space tend to escape towards infinity with an exponential escape rate. The number 
of particles $N(\tau)$ that remain inside the initial domain at instant $\tau$ decays exponentially with time as 
$N(\tau)\sim N(0) e^{-\gamma \tau}$, being $\gamma$ the so called {\sl escape rate}.

The leading eigenvalue of $\hat P^\tau$ dominates the decay and determines the escape rate. The rest of the spectrum 
({\it resonances}) describe further details of the dynamics and give other important time scales besides the one
associated with the escape rate. Thus the spectrum of the Frobenius-Perron operator $\hat P^\tau$ provides a way to
describe the evolution of a set of trajectories characterized by a density $\rho_\tau(X)$. 

The spectral theory of $\hat P^\tau$ developed from the works of Koopman \cite{koopman1931} and von Neumann \cite{neumann1932}. 
It assumes that the evolution operator acts on a functional space of square integrable densities ${{\cal {L}}^2}$. In that 
case the evolution is unitary and the eigenvalues of $\hat P^\tau$ are located on the unit circle. The spectrum of a chaotic 
system presents continuous components on the unit circle which describe the correlation function decay for these systems. New 
methods have been developed to obtain the eigenvalues of $\hat P^\tau$ outside the unit circle and to analyze the escape 
process or relaxation dynamics \cite{pollicot1985,ruelle1986}. These methods, which are valid for systems where all 
the periodic orbits are unstable, have been successfully used in chaotic systems and the study of transport 
phenomena among other applications \cite{gaspard1992,gaspard1998,gaspard1995,alonso1996,alonso1993}. We shall apply 
them to the study of the escape dynamics of photons from the two black holes.

One of the methods developed to obtain the spectrum computes the trace of the Frobenius-Perron operator. Here we shall 
outline this procedure and refer the reader to the specialized literature on the subject 
\cite{pollicot1985,ruelle1986,isola1988,gaspard1998,gaspard1995} for further details.

The trace of $\hat P^\tau$ can be formally written as
\begin{equation}
\label{trace}
Tr \hat P ^\tau = \int dX \, \delta (X-\Phi^\tau X).
\end{equation}
The contributions to the trace are due to the fixed points of the flow, which are given by the condition
\begin{equation}
\label{fixedpoint}
X=\Phi^\tau X,
\end{equation}
at some $\tau$. In general this equation presents two types of solutions: {\it stationary points} that satisfy 
(\ref{fixedpoint}) for all $\tau$, and {\it periodic orbits} which satisfy the fix-point condition at a discrete 
set of values, given by $\tau=rT_p$, with $r=1,2,\cdots$. We assume that the solutions of (\ref{fixedpoint}) are 
isolated, and therefore each stationary point or periodic orbit is locally unique. A periodic orbit may belong to a 
continuous family provide that there exists some continuous symmetry or some constant of motion. When
this occurs the integration domain in (\ref{trace}) must be reduced to a less dimensional space until the periodic 
orbit is completely isolated. In time-independent Hamiltonian systems the periodic orbits are rarely isolated, 
indeed they tend to form continuous families as the energy changes. Hence the phase-space must be reduced 
considering the energy conservation ${\cal H}=E$. Then the trace (\ref{trace}) is formally given by
\begin{equation}
\label{trace2}
Tr_{E} \hat P^\tau = \int_{E} d^{2f-1}x \,\delta(X-\phi_E^\tau x),
\end{equation}
where $\phi_E^\tau$ denotes the flow on the energy shell \cite{cvitanovic1991,gaspard1998}. If any additional symmetry is 
present in the system then further reductions are required. In the two black system we are considering the dynamics on 
the energy shell and we have removed the axial symmetry by choosing a particular meridian plane. Therefore all the periodic 
orbits of the flow can be treated as isolated.

The trace (\ref{trace2}) can be written in terms of the unstable periodic orbits of the flow and their repetitions 
\cite{artuso1990,cvitanovic1991}. In order to do so, the integral in (\ref{trace2}) is resolved considering at each point 
a coordinate system with one axis fixed along the periodic orbit. The integral over the coordinate along the orbit is
trivial, being related to the period of the orbit. The integration over the transverse coordinates takes into account
the stability of the periodic orbit. The result for the trace is \cite{cvitanovic1991}
\begin{equation}
\label{trace3}
Tr \hat P^\tau =\sum_{p=p.p.o.} \sum_{r=1}^{\infty} T_p \frac{\delta (\tau-rT_p)}{|{\hbox{det}}({\sf 1}-{\sf m}_p^r)|} 
\hspace*{0.4cm}(\tau>0),
\end{equation}
where the sums are over the {\it primitive periodic orbits (p.p.o.)} and their $r$ repetitions. 
The matrix ${\sf m}_p$ is derived from the fundamental matrix ${\sf M}(X_0,T_p)$ once the perturbations along the orbit 
and the perturbations perpendicular to the energy shell have been removed. Hence in a two-dimensional flow the matrix 
${\sf m}_p$ only contains the stretching factors whose modulus is different from one. These are precisely the 
ones that are related to the stable and unstable manifolds of the isolated unstable periodic orbit.

Formally the Laplace transform of the Frobenius-Perron operator gives the resolvent of the Liouville operator defined 
in (\ref{liouville}), which using (\ref{trace2}) can be written as
\begin{equation}
Tr \frac{1}{\sigma - \hat L}= \int_0^{\infty} d\tau \, e^{-\sigma \tau} Tr \hat P= \frac{\partial}{\partial \sigma} 
\ln Z(\sigma), 
\end{equation}
where $Z(\sigma)$ is the so called Selberg-Smale Zeta function \cite{smale1980}, given by
\begin{equation}
Z(\sigma)\equiv \exp \left( - \sum_{p=p.p.o.} \sum_{r=1}^{\infty} \frac{1}{r} 
\frac{e^{-\sigma r T_p}}{|{\hbox{det}}({\sf 1}-{\sf m}_p^r)|} \right).
\end{equation}
Denoting by $\Lambda_p$ and $\Lambda_p^{-1}$ the eigenvalues of ${\sf m}_p$, the zeta function can be expressed as
\begin{equation}\label{zf}
Z(\sigma)=\prod_{p=p.p.o} \prod_{k=0}^{\infty} \left( 1- \frac{e^{-\sigma T_p}}
{ | \Lambda_p |\, \Lambda_p ^k}\right)^{k+1}
\end{equation}
where $ \Lambda_p > 1$ since all the periodic orbits are unstable in the two black-hole system. The spectrum of the system
can be determined from the zeroes of the function $Z(\sigma)$, which is usually expressed as a product of the inverse
{\it Ruelle} $\zeta$-functions,
\begin{equation}
Z(\sigma)= \zeta_0^{-1}(\sigma) \zeta_1^{-2}(\sigma)\zeta_2^{-3}(\sigma) \cdots
\end{equation}
with the definition
\begin{equation}
\zeta_k^{-1}(\sigma)= \prod_{p=p.p.o} 
\left(
1-\frac{e^{-\sigma T_p}}{|\Lambda_p|\Lambda_p^k}
\right).
\end{equation}

The relaxation times $\tau_i$ associated with the dynamics are given by the poles of the resolvent of the Liouvillian, as 
$Re \, \sigma_i=\tau_i^{-1}$, or equivalently, by the zeroes of the zeta function or the zeroes of the inverse of the 
Ruelle $\zeta$-functions. The leading part of the spectrum is given by the poles of the first Ruelle 
$\zeta$-function $\zeta_0(\sigma)$ (or zeroes of $\zeta_0^{-1}$). So we are mainly interested in the part  
of the spectrum that controls the long-time behavior of the dynamics.

The set of unstable periodic orbits can be used to describe in fine detail the escape dynamics of photons from the black 
holes. The number of periodic orbits in the fractal repeller grows exponentially with the period, with a rate of proliferation 
that is given by the topological entropy $h_{top}$. The amount of chaos in the system is given by the Kolmogorov-Sinai 
entropy per unit time $h_{KS}$, which measures the minimal data accumulation rate required to reconstruct a trajectory on 
the repeller without ambiguity. The repeller has a mean Lyapunov exponent $\lambda$. In a Poincare section that is transverse 
to the flow on the repeller, the fractal repeller defines another fractal which is characterized by the partial generalized 
fractal dimensions. Two representative partial dimensions are the Hausdorff dimension $d_H$, which characterizes the bulkiness 
of the repeller in phase space, and partial information dimension $d_I$ that can be obtained from the Kolmogorov-Sinai entropy 
and the mean Lyapunov exponent as $d_I= h_{KS}/\lambda$ \cite{gaspard1989}.

The different quantities that characterize the repeller can be obtained from the so called topological pressure $P(\beta)$
\cite{ruelle1978}, which can be determined from the periodic orbits that form part of the repeller as the leading zero of a 
Ruelle $\zeta$-function
\begin{equation}
\zeta_\beta^{-1}(\sigma) \equiv \prod_{p=p.p.o} 
\left(
1-\frac{e^{-\sigma T_p}}{|\Lambda_p|^{\beta}}
\right).
\end{equation}

The mean Lyapunov exponent for the trajectories on the fractal repeller is $\lambda=-dP(1)/d\beta$, the escape rate is 
$\gamma=-P(1)$ and the topological entropy is given by $h_{top}=P(0)$. In two-dimensional Hamiltonian systems the 
Kolmogorov-Sinai entropy follows from $h_{KS}=\lambda-\gamma$ \cite{pesin}. The partial Hausdorff dimension is obtained 
from $P(d_H)=0$ and the information dimension is given by  $d_I=1-\gamma / \lambda$.

\section{Periodic orbits and the repeller in the meridian plane $(L_z=0)$}

From now on we shall focus on the analysis of the escape dynamics of photons from the black holes in the meridian plane.
As we have just mentioned, there are some important objects that can be used to describe this dynamics.
First we need to identify the set of periodic orbits  and their corresponding primitive periods. To analyze the linear 
stability of each unstable periodic orbit we must integrate its fundamental matrix in time up to one primitive period.
The eigenvalues of the resulting matrix give the stretching factors associated with the orbit.
Then, once we know the periods and the stretching factors of the unstable periodic orbits, the main quantities of 
chaos that characterize the dynamics can be obtained  following the methods described in the previous section.

The restriction on the motion to the meridian plane $L_z=0$ simplifies the set of equations to be integrated. Without 
loss of generality we can consider that the photons move in the $(x,z)-$plane. Then the equations of motion (\ref{eqnmotion}) 
with respect to the affine parameter $\mu$ reduce to
\begin{equation}
\dot \alpha=-U^{-2}p_\alpha
\hspace*{0.5cm}\hspace*{0.5cm}
\dot p_\alpha=-2U\partial_\alpha U
\hspace*{0.5cm}(\alpha=x,z),
\end{equation}
being the equations (\ref{eqnmotion2}) with respect to the killing time $t$,
\begin{equation}
\alpha^\prime=-U^{-4}p_\alpha
\hspace*{0.5cm}\hspace*{0.5cm}
p_\alpha^\prime=-2U^{-1}\partial_\alpha U
\hspace*{0.5cm}(\alpha=x,z).
\end{equation}
These equations of motion and the equations associated with the stability analysis were simultaneously integrated using a 
Runge-Kutta-Fehlberg method. The search for periodic orbits was performed using the Newton-Raphson method.
As we just mentioned, to compute the stretching factors of the set of unstable periodic orbits, each orbit was numerically 
integrated together with its corresponding fundamental matrix up to one primitive period of the orbit. 
Considering that the dynamics is highly unstable, it is important to compute the unstable periodic orbit with high 
accuracy in order to obtain the stability factors with good precision. 
In our numerical results the required high accuracy in the period of the orbits and the stretching factors was achieved
by means of a Mathematica code in which a very high precision was demanded.

\begin{figure}[h]
\centerline{\includegraphics[width=0.93\textwidth]{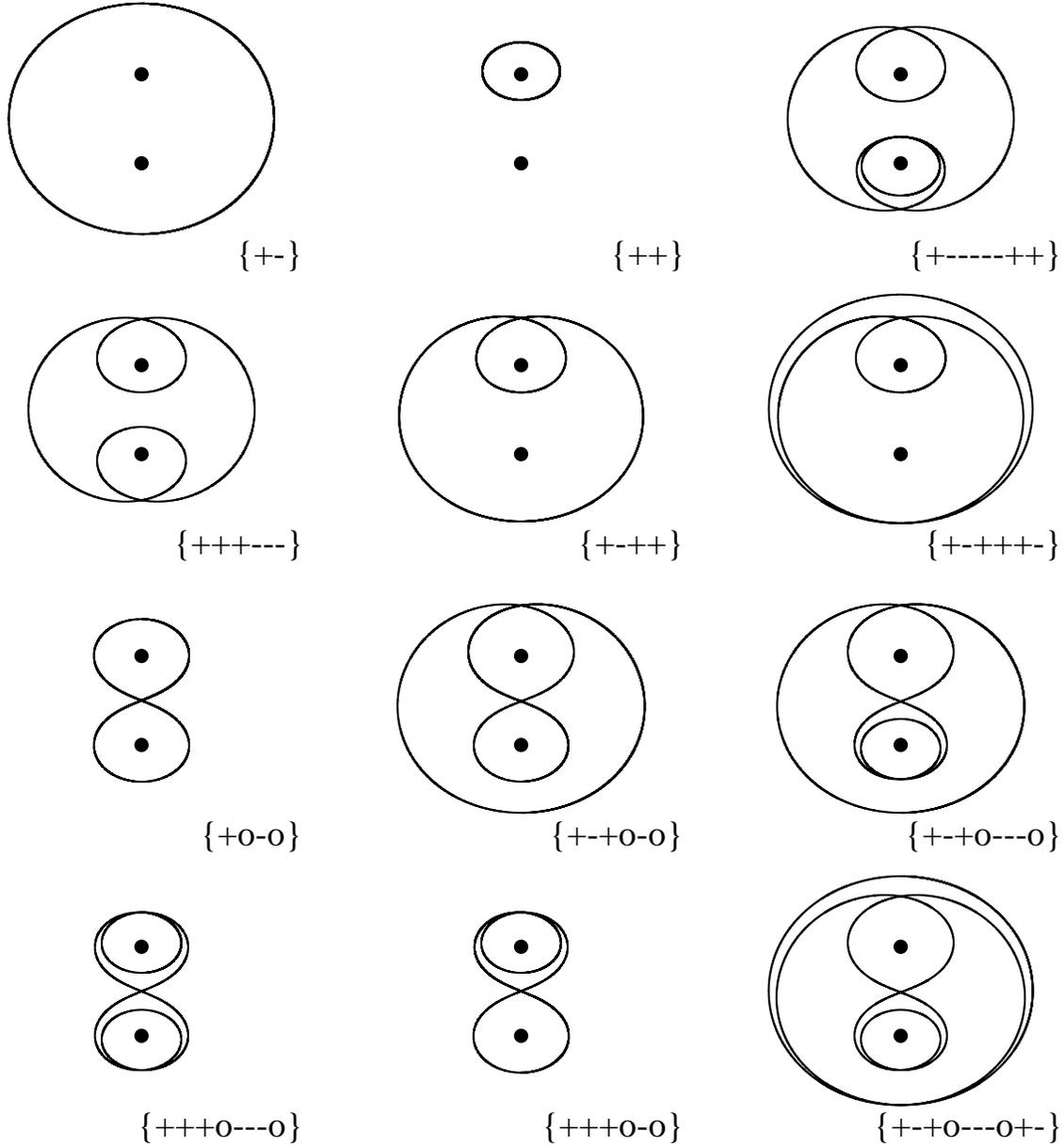}}
\vspace*{0.8cm}
\caption{Some of the unstable periodic orbits in a system with two black holes located at $z_{bh}=\pm 1$ and masses $M_1=M_2=1$.
The symbolic coding that labels each orbit is indicated. Due to the symmetry of the system with respect to equatorial line 
(the $x$-axis), the change of sign of the coordinate $z$ in each orbit on the last two columns gives an additional periodic 
orbit. The symbolic coding for these {\sl inverse} orbits is obtained exchanging the symbols "+" and "-" in the orbits depicted 
in these columns. For instance, the inverse orbit of $\{++\}$ has symbolic coding $\{--\}$, the inverse of $\{+-++\}$ is 
$\{-+--\}$, the inverse of $\{+-+\circ-\circ\}$ is $\{-+-\circ+\circ\}$ and so on.}
\label{po1}
\end{figure}

The periodic orbits in the system can be classified according to a symbolic coding. In this work, 
taking as reference the symmetry axis that contains the two black holes, in our coordinate system the $z$-axis, we 
introduce a series of symbols to label these orbits. An orbit can cross the $z$-axis either above the black-hole located 
at $z=z_{bh}$, below the black-hole located at $z=-z_{bh}$ or in between the two black-holes. 
We label each crossing that occurs at $z>z_{bh}$ with the  symbol ``$+$'', to indicate a crossing at $z<-z_{bh}$ we use the 
symbol ``$-$''. When an orbit crosses the $z$-axis twice at the same point in between the two black holes, drawing an 
``$\times$'' on the axis, we use the symbol ``$\circ$''. Any other crossing in between the two black holes is indicated with the 
symbol ``$+$'' or the symbol ``$-$'' depending on the sign of the $z$ coordinate at the crossing point, $z>0\,(+)$ or $z<0\,(-)$.
Figure (\ref{po1}) displays some of the unstable periodic orbits numerically found and their corresponding symbolic
coding. The primitive period of each orbit is given in Table (\ref{table1}).

In \cite{cornishgibbons} a different coding to label the periodic orbits was introduced. This coding was used to compute the
topological entropy per unit symbol. We remark that one of our goals in this work is to compute the topological entropy per 
unit time.

\begin{table}
\begin{tabular}{lrrr}\hline
Symbolic coding & \hspace*{0.5cm}Period $T_p\,(\mu)$ &  \hspace*{0.5cm}Period $T_p\,(t)$ & 
\hspace*{0.5cm}Stretching factor $\Lambda_p$\hspace*{0.3cm}\\ \hline 
$\{++\}$, $\{--\}$ & \hspace*{0.5cm}4.1180374 & \hspace*{0.5cm}37.9255026 & \hspace*{0.5cm}99.1244899\hspace*{0.3cm}\\
$\{+-\}$ & \hspace*{0.5cm}15.292055 & \hspace*{0.5cm}52.7275448 & \hspace*{0.5cm}215.721472\hspace*{0.3cm} \\
$\{+\circ-\circ\}$ & \hspace*{0.5cm}10.608412 & \hspace*{0.5cm}79.8770864 & \hspace*{0.5cm}16996.3144\hspace*{0.3cm} \\
$\{+-++\}$, $\{-+--\}$ & \hspace*{0.5cm}18.900782 & \hspace*{0.5cm}91.7239060 & \hspace*{0.5cm}30044.8517\hspace*{0.3cm}\\
$\{+++\circ-\circ\}$,  $\{---\circ+\circ\}$ & \hspace*{0.5cm}14.739760 &\hspace*{0.5cm}117.837331 &  
\hspace*{0.5cm}$1.70990814 \times 10^6$\hspace*{0.3cm} \\
$\{+++---\}$& \hspace*{0.5cm}22.635603& \hspace*{0.5cm}130.594585 &  \hspace*{0.5cm}$3.97082132 \times 10^6 $\hspace*{0.3cm}\\
$\{+-+\circ-\circ\}$,  $\{-+-\circ+\circ\}$ & \hspace*{0.5cm}25.410381 & \hspace*{0.5cm}133.342257 & 
\hspace*{0.5cm}$4.57097810 \times 10^6$\hspace*{0.3cm} \\
$\{+-+++-\}$,  $\{-+---+\}$ & \hspace*{0.5cm}34.185532 & \hspace*{0.5cm}144.455986 &  \hspace*{0.5cm}$6.49738275 \times 10^6$
\hspace*{0.3cm} \\
$\{+++\circ---\circ\}$ & \hspace*{0.5cm}18.871204 & \hspace*{0.5cm}155.798106 &  \hspace*{0.5cm}$1.72085241 \times 10^8$
\hspace*{0.3cm} \\
$\{+-----++\}$,  $\{-+++++--\}$ & \hspace*{0.5cm}26.748625 & \hspace*{0.5cm}168.528893 &  \hspace*{0.5cm}$ 3.95770189 
\times 10^8$\hspace*{0.5cm} \\
$\{+-+\circ---\circ\}$, $\{-+-\circ+++\circ\}$ & 
\hspace*{0.5cm}29.542461 & \hspace*{0.5cm}171.300173 &  \hspace*{0.5cm}$4.59225358 \times 10^8$\hspace*{0.5cm} \\
$\{+-+\circ---\circ+-\}$,  $\{-+-\circ+++\circ-+\}$ & 
\hspace*{0.5cm}44.828729 & \hspace*{0.5cm}224.030938 &  \hspace*{0.5cm}$ 9.92347816 \times 10^{10}$\hspace*{0.5cm}\\ \hline
\end{tabular}
\caption{The periods and the stretching factors of the unstable periodic orbits depicted in figure (\ref{po1}). The periods
corresponding to both the affine parameter $\mu$ and the killing time $t$ are indicated.}
\label{table1}
\end{table}

In Table (\ref{table1}) are listed the non-trivial stretching factors of the periodic orbits depicted in figure (\ref{po1}). 
We recall that these are the stretching factors that are related to the directions transverse to the periodic orbit.
The trajectories $\{++\} (\{--\})$, $\{+-\}$ and $\{+\circ-+\}$ have the lowest stability eigenvalues. The rest of the orbits 
are much more unstable as they have larger stretching factors. In general the more stable periodic orbits should dominate 
the asymptotic escape of particles from the two black-hole configuration.

In the two black-hole system the set of unstable periodic orbits are the only critical elements of the flow in the fractal 
repeller that determines the escape dynamics of photons. This distinguished subset of solutions plays a crucial role as the 
averages of the dynamical quantities can be obtained in terms of periodic orbits, provide that it is dense on the repeller 
\cite{gaspard1998}.

The dynamics in the proximity of the invariant set that defines the repeller can be analyzed considering the time interval 
that it takes to an orbit to escape from a finite spatial domain ${\cal U}$, when evolves both forwards and backwards in time. 
To numerically study the escape dynamics we introduce the forward and backward escape time functions \cite{gaspard1998}, 
which are defined by: 
\begin{eqnarray}
T_{\cal U}^+(x)&=&Max \lbrace T>0\,;\,\, \Phi^{\mu} x \in {\cal U}\,, \,\, \forall \mu \in [0, T[\,, \,\, \forall x 
\in {\cal U} \rbrace \\
T_{\cal U}^-(x)&=&Min \lbrace T<0\,;\,\, \Phi^{\mu} x \in {\cal U}\,, \,\, \forall \mu \in \,]T, 0]\,, \,\, \forall x 
\in {\cal U} \rbrace
\end{eqnarray} 

The escape time functions provide an useful method to construct the repeller \cite{burghardt1994}. The ensemble of initial 
conditions whose escape time coordinate is larger than a given value $T$ contains all the trajectories that still remain inside 
the domain $\cal U$ at time $T$. In the long time limit this set contains the trapped trajectories of the repeller and its 
stable manifolds in $\cal U$. As time evolves this set splits into smaller pieces and eventually becomes a fractal set. 
Similarly it is possible to find the set of initial conditions that have already arrived in the domain $\cal U$ before 
some time $-T$ in the past. The repeller is given by the intersection of these two sets. 

\begin{figure}[h]
\vspace*{1cm}
\centerline{\includegraphics[width=1.0\textwidth]{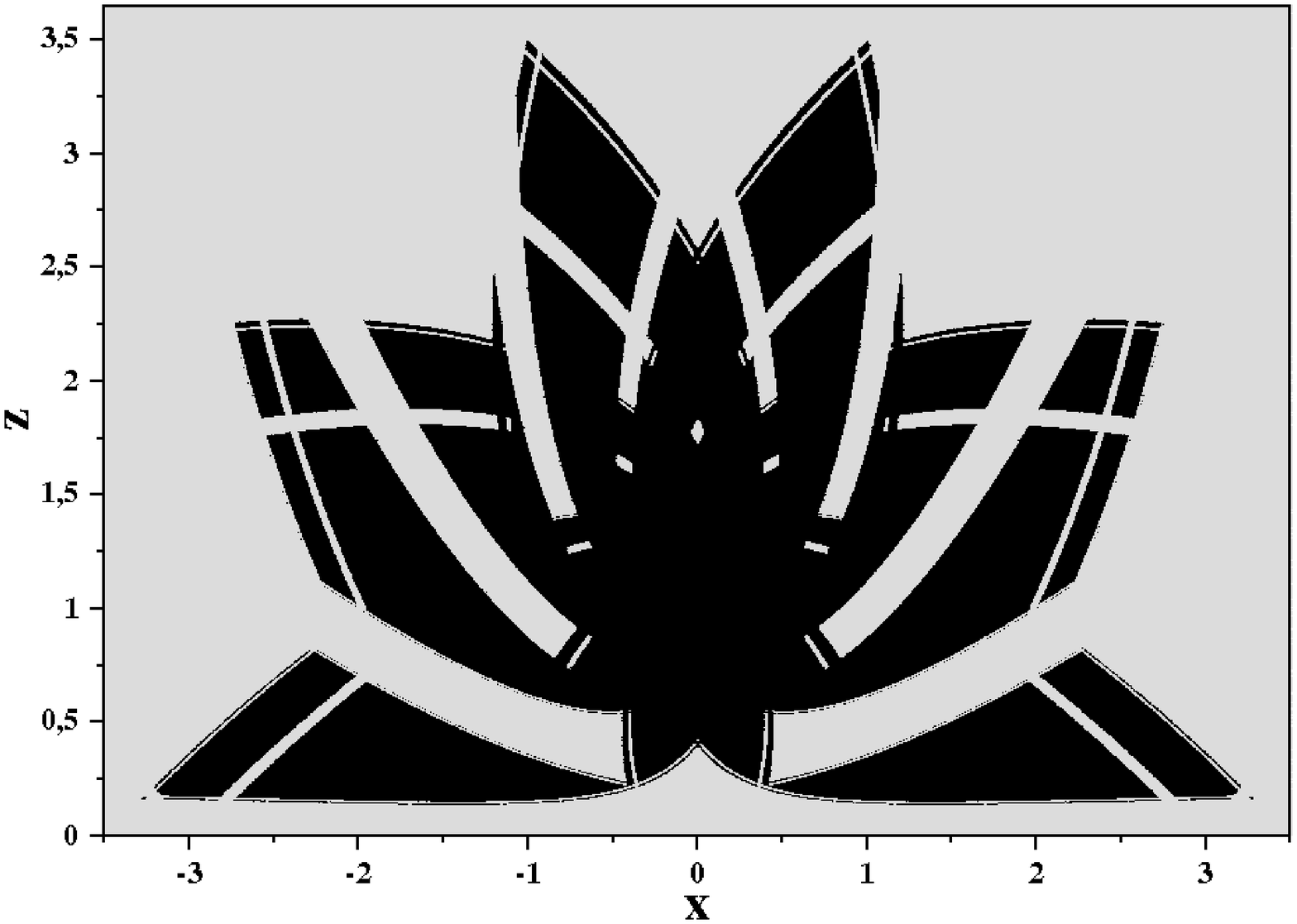}}
\caption{Outcomes for photons in the two-black-hole field. The black dots are the initial positions of the trajectories which
fall into any of the two black holes. The gray dots localize the initial positions of the photons that escape towards infinity.
The evolution according to the equations of motion (\ref{eqnmotion}) was considered, both backwards and forwards in time.
The initial positions $(x,z)$ were chosen at random in the finite region depicted, and the initial directions were randomly
selected with an angle with respect to the $x$-axis equals to $0$, $\pi/5$, $2\pi/5$, $3\pi/5$ or $4\pi/5$.}
\label{repel1}
\end{figure}

\begin{figure}[h]
\centerline{\includegraphics[width=1.0\textwidth]{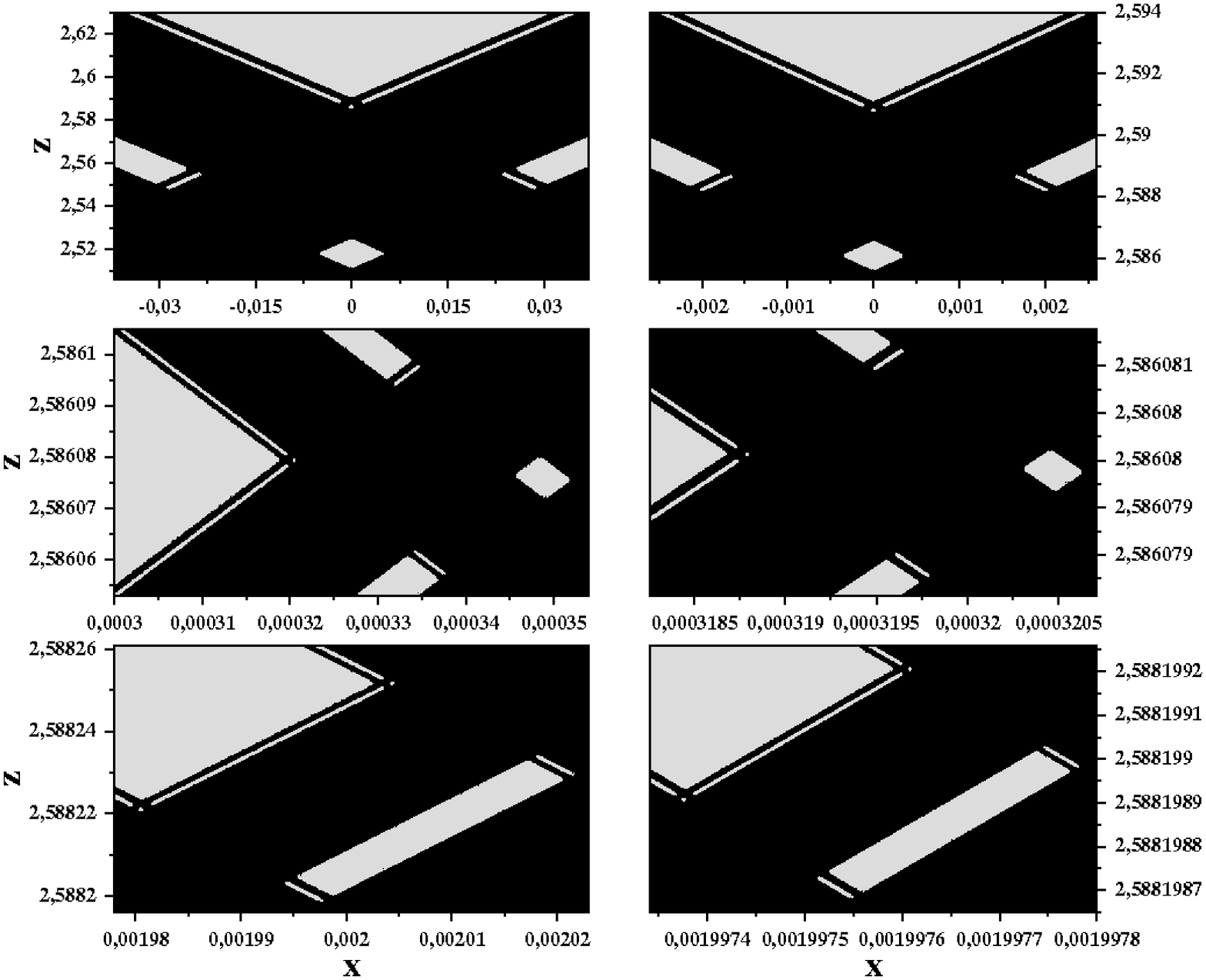}}
\caption{Several blown up regions of figure (\ref{repel1}) showing the fractal structure of the repeller.}
\label{repel2}
\end{figure}

The compound escape function $T_{\cal U}^+(x)+|T_{\cal U}^-(x)|$ is very useful to illustrate the structure of the repeller.
To numerically compute this function we analyze the evolution, both backwards and forwards in time, according to the 
equations of motion (\ref{eqnmotion}), of a large ensemble of photons initially distributed at random in a finite region of 
space and certain fixed initial directions. 
In order to distinguish the possible outcomes of the photons in the two black-hole field, we color code the initial position 
black if the corresponding photon falls into any of the two black holes and gray if the photon escapes towards infinity.
Figure (\ref{repel1}) displays a portrait of the basins of attraction associated with these two possible outcomes in 
the plane $(x,z)$ of initial positions, for an ensemble of photons with initial directions randomly fixed between 
five selected values for the angle with respect to the $x$-axis. It is well known that for chaotic systems the interweaving 
of outcomes leads to fractal basin boundaries; and in the two black-hole system, as expected, we can see them explicitly.
The blown up regions depicted in the different panels of figure (\ref{repel2}) exemplify the repeated fractal structure in 
the basin boundaries that characterize the hyperbolic repeller.
In the following sections we compute different characteristic quantities of the dynamics related to this fractal repeller.

\section{Escape rate in the meridian plane and other characteristic quantities of Chaos}

Using the primitive periods $T_p$ and stretching factors $\Lambda_p$ of the set of unstable periodic orbits we can compute
the escape rate of the system as well as other characteristic quantities of the dynamics related to the fractal repeller, 
following the methods described in Section IV. To this aim we consider the cycle expansion method 
\cite{artuso1990,cvitanovic1991} which has been successfully applied to other hyperbolic systems \cite{gaspard1992}. 
Due to the high instability of higher order periodic orbits (longer symbolic coding) only three main periodic orbits, 
$\{++\} (\{--\})$, $\{+-\}$ and $\{+\circ-+\}$, contribute significantly to the escape rate. 
Indeed they give an acceptable escape rate and the different quantities of chaos.

\begin{figure}[h]
\centerline{\includegraphics[width=0.7\textwidth]{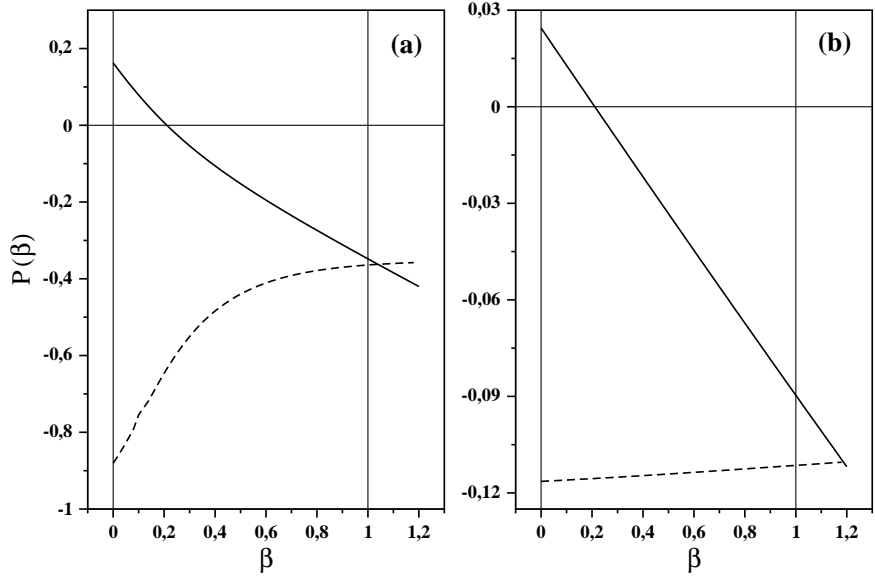}}
\caption{The pressure function (solid line) obtained considering the periodic orbits that include up to six elements 
in the symbolic code, see figure (\ref{po1}). The dashed line is the first derivative of this pressure function. The 
panel $(a)$ is the pressure associated with the affine parameter, and $(b)$ the pressure corresponding to the killing 
time
\vspace*{0.5cm}
.}
\label{pressure}
\end{figure}

Figure (\ref{pressure}) shows the pressure function obtained from these three periodic orbits. The main quantities 
of chaos obtained from this pressure function and its first derivative are listed in Table (\ref{table2}).
\begin{table}
\begin{tabular}{|c||c|c|c|c|c|c|} \hline
    & $\lambda$ &  $\gamma$ &  $h_{KS}$ &  $h_{top}$ &  $d_{H}$ &  $d_{I}$ \\ \hline\hline
\hspace*{0.3cm} Affine parameter $\mu$ \hspace*{0.3cm} &
\hspace*{0.3cm}  0.364  \hspace*{0.3cm} &\hspace*{0.3cm}  0.348  \hspace*{0.3cm}&  
\hspace*{0.3cm} 0.016  \hspace*{0.3cm}&\hspace*{0.3cm}  0.162  \hspace*{0.3cm}&
\hspace*{0.3cm} 0.221  \hspace*{0.3cm}&\hspace*{0.3cm}  0.045 \hspace*{0.3cm} \\ \hline
\hspace*{0.3cm} Killing time $t$ \hspace*{0.3cm} &
\hspace*{0.3cm}  0.111  \hspace*{0.3cm} &\hspace*{0.3cm}  0.089  \hspace*{0.3cm}&
\hspace*{0.3cm} 0.022  \hspace*{0.3cm}&\hspace*{0.3cm}  0.024  \hspace*{0.3cm}&
\hspace*{0.3cm} 0.221  \hspace*{0.3cm}&\hspace*{0.3cm}  0.196 \hspace*{0.3cm} \\ \hline
\end{tabular}
\caption{Characteristic quantities of chaos related to the fractal repeller associated with the unstable periodic orbits 
given in figure (\ref{po1}). $\lambda$ is the mean Lyapunov exponent of the repeller, $\gamma$ the escape rate, $h_{KS}$ 
the Kolmogorov-Sinai entropy, $h_{top}$ the topological entropy, $d_H$ the Hausdorff dimension and $d_I$ the partial
information dimension.}
\label{table2}
\end{table}
In particular, the escape rate which is equal to $\gamma_\mu=0.348$ for the affine parameter and $\gamma_t=0.089$ for the
killing time. From the analysis of the linear stability of the set of unstable periodic orbits, it could be assumed that 
the escape dynamics of photons from the two black holes is mainly controlled by the outermost periodic orbit $\{+-\}$. 
If only this orbit is considered it follows an escape rate
\begin{equation}
\gamma_\mu^{\{+-\}}=\frac{\ln |\Lambda_{+-}|}{T_{+-}(\mu)}=0.351
\hspace*{2cm}
\gamma_t^{\{+-\}}=\frac{\ln |\Lambda_{+-}|}{T_{+-}(t)}=0.102
\end{equation}
which is close to the escape rate obtained from the pressure function. Thus, we can conclude that the orbit $\{+-\}$ 
dominates the effective escape of photons from the two black-hole configuration, and even provides a fair estimate of 
the escape rate. 

In the following section we contrast the escape rate value derived from the analysis of the linear stability of the 
periodic orbits with the escape rate obtained from numerical simulations that consider the time evolution of an  
statistical ensemble of photons. 

\subsection{Numerical simulations with statistical ensembles}

We consider an ensemble of photons with initial positions uniformly distributed along the $z$-axis and initial velocity 
parallel to the x-axis. This together with the energy conservation completely defines the ensemble. Once all the initial 
conditions have been fixed the time evolution of each photon of the ensemble is numerically integrated. Some of the 
photons fall into one of the two black holes, while others escape to infinity. We are precisely interested in the second 
subset of trajectories. 
To compute the escape rate we consider a circle of control that covers the two black holes and  which is larger than the 
outermost periodic orbit $\{+-\}$. Hence all the photons that cross out this circle escape towards infinity and do not come 
back to the proximity of the black holes.

Let us denote by $N(\tau)$, with $\tau$ the affine parameter $\mu$ or the killing time $t$, the number of photons that 
remain inside the circle of control at time $\tau$ (this includes the photons that fall into the black-holes). This 
function decreases with $\tau$ until it reaches an asymptotic value $N(\infty)$, which gives the number of photons 
that fall into the black holes. The escape rate $\gamma$ is defined by the exponential decrease to zero of the function 
$[N(\tau)-N(\infty)] \approx [N(0)-N(\infty)]e^{-\gamma\tau}$ as $\tau\to\infty$.

\begin{figure}[h]
\centerline{\includegraphics[width=0.8\textwidth]{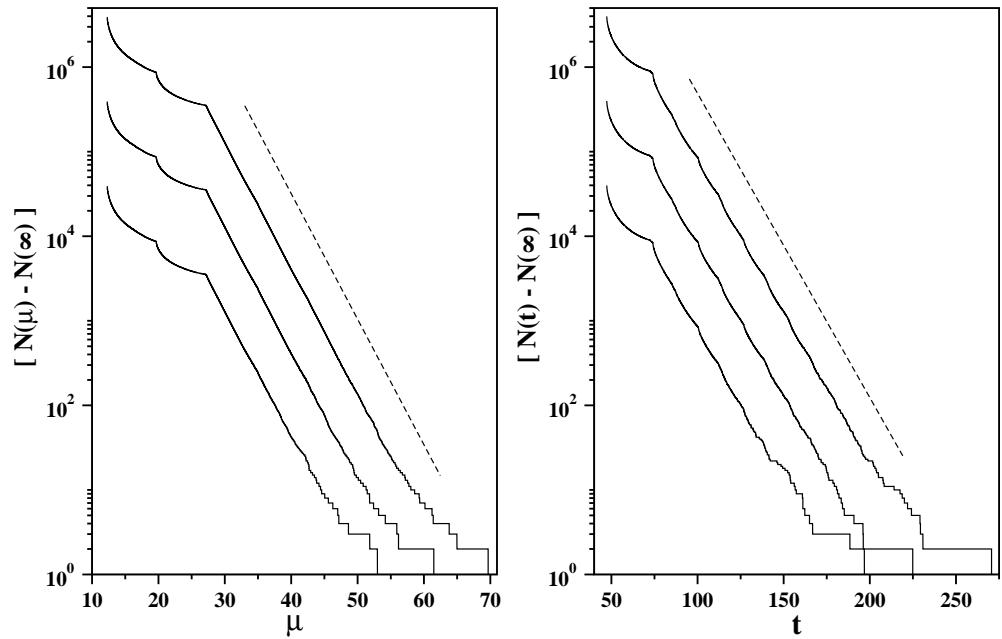}}
\caption{The evolution of $[N(\mu)-N(\infty)]\,\,(\,[N(t)-N(\infty)]\,)$ with the affine parameter (killing time) for 
three ensembles with an increasing number of photons, $5\times10^5$, $5\times10^6$ and $5\times10^7$. The dashed lines give 
the best linear fitting for the ensemble with the highest number of photons.}
\label{nt}
\end{figure}

\begin{figure}[h]
\centerline{\includegraphics[width=0.8\textwidth]{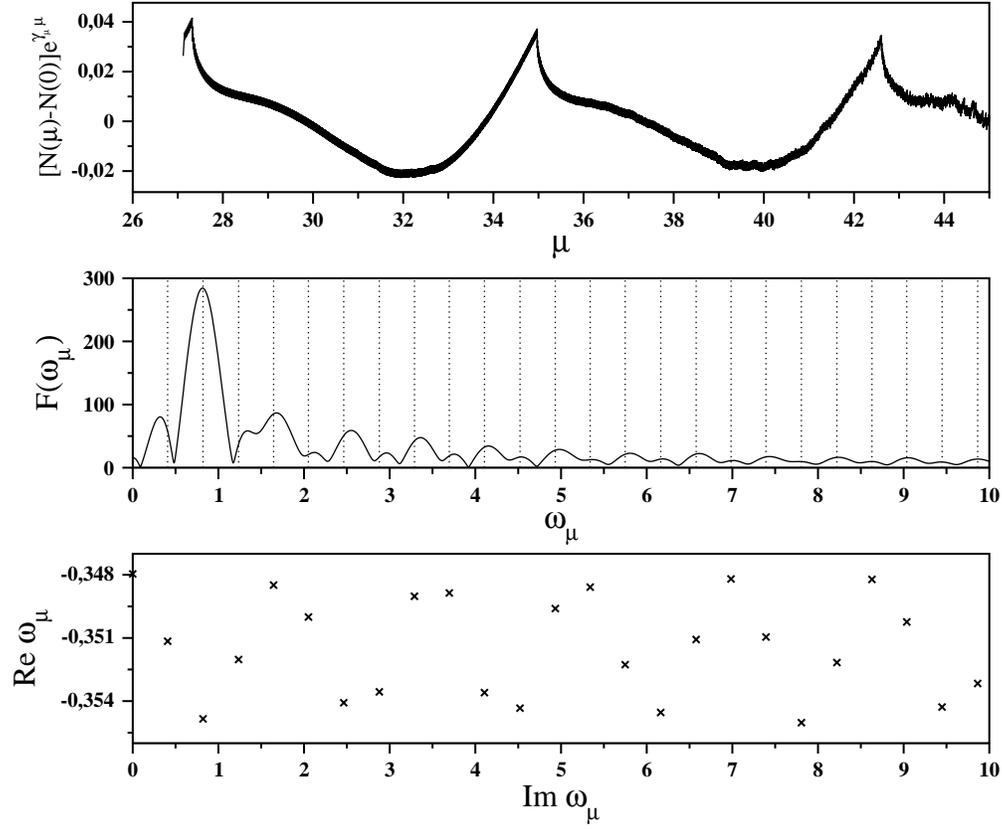}}
\caption{In the upper panel the oscillatory component of the function $[N(\mu)-N(\infty)]$ depicted in the left panel of
figure (\ref{nt}), once the main exponential decay has been subtracted. In the middle panel its power spectrum, being the 
vertical dotted lines the imaginary part of the resonances given in the lower panel.} 
\label{nt2a}
\end{figure}

\begin{figure}[h]
\centerline{\includegraphics[width=0.8\textwidth]{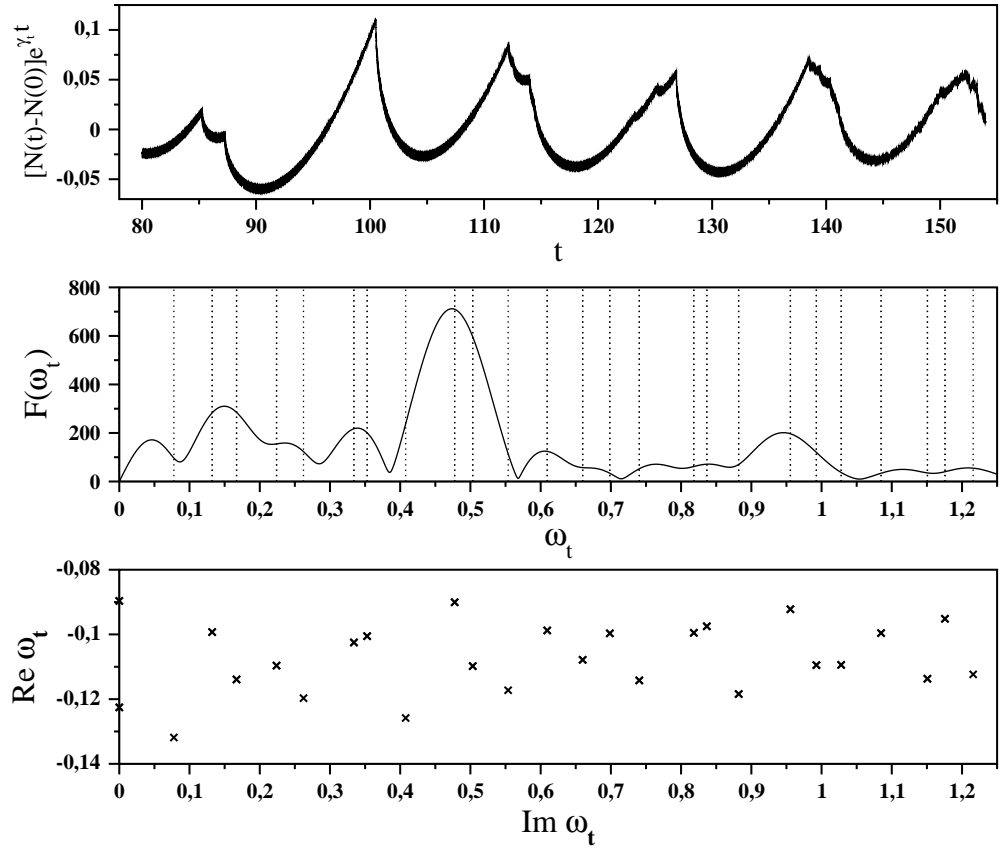}}
\caption{The same as figure (\ref{nt2a}), but for the function $[N(t)-N(\infty)]\,)$ depicted in the right panel of 
figure (\ref{nt}).}
\label{nt2b}
\end{figure}

Figure (\ref{nt}) shows the results obtained from different simulations that include an increasing number of particles 
in the statistical ensemble. The larger is the number of particles in the ensemble the longer persists the exponential 
decrease with $\tau$. 
From the slope of these curves we extract a numerical escape rate $\gamma_{\mu}^{(num)}=0.344$ for the affine parameter
and $\gamma_{t}^{(num)}=0.083$ for the killing time. The agreement with the escape rate obtained from the analysis of the 
linear stability of the unstable periodic orbits is excellent, see Table (\ref{table2}). 
This result confirms our expectation with respect to the subset of periodic orbits that plays the main role in the escape 
dynamics. The small oscillatory component in the evolution of the function $[N(\tau)-N(\infty)]$ can be explained by 
the lower resonances of the system. This point has been emphasized in \cite{gaspard1992}. To further illustrate this
aspect we have computed the power spectra of the numerical data $[N(\mu)-N(\infty)]e^{\gamma_{\mu}^{(num)}\mu}$ for the affine 
parameter, and $[N(t)-N(\infty)]e^{\gamma_{t}^{(num)}t}$ for the killing time. The figures (\ref{nt2a}) and (\ref{nt2b}) show 
a peak structure associated with the oscillations of both functions once the main exponential decay has been subtracted. The 
positions of the peaks is compared with the locations of the imaginary part of the lower resonances computed from the zeta 
function taking orbits that include up to six elements in the symbolic code. The resonance distribution explains the peak 
structure observed in the power spectrum.

\begin{figure}[h]
\centerline{\includegraphics[width=0.73\textwidth]{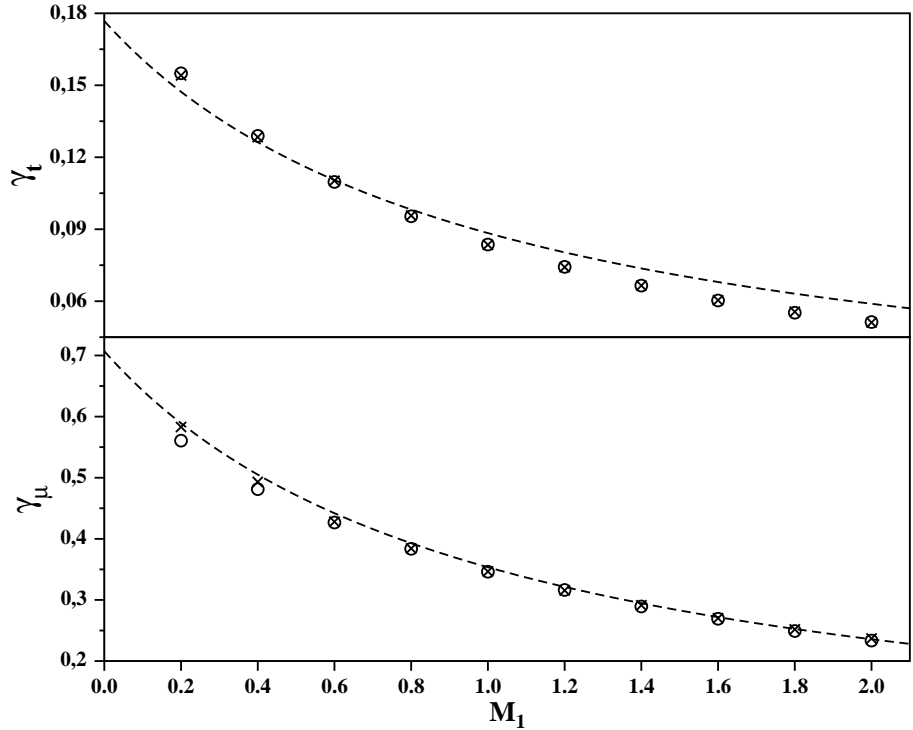}}
\caption{The escape rate $\gamma_\mu\,\,(\,\gamma_t\,)$ from the two black holes located at $z_{bh}=\pm 1$ for different 
values of the mass $M_1$ ($M_2=1$). The data were obtained considering ensembles of photons with initial velocity parallel 
to the x-axis and initial positions uniformly distributed along an interval on the $z$-axis. The circles correspond to an 
ensemble initially distributed in the interval $[1.1,4]$ and the crosses to an ensemble initially distributed in the interval 
$[-4,-1.1]$. The dashed lines are the analytic estimate of the escape rate from a perturbed black hole, 
$\gamma_\mu^{(0)}=1/\sqrt{2}(M_1+M_2)$ and $\gamma_t^{(0)}=1/4\sqrt{2}(M_1+M_2)$, see Section VII.}
\label{nt2}
\end{figure}

\begin{figure}[h]
\centerline{\includegraphics[width=0.73\textwidth]{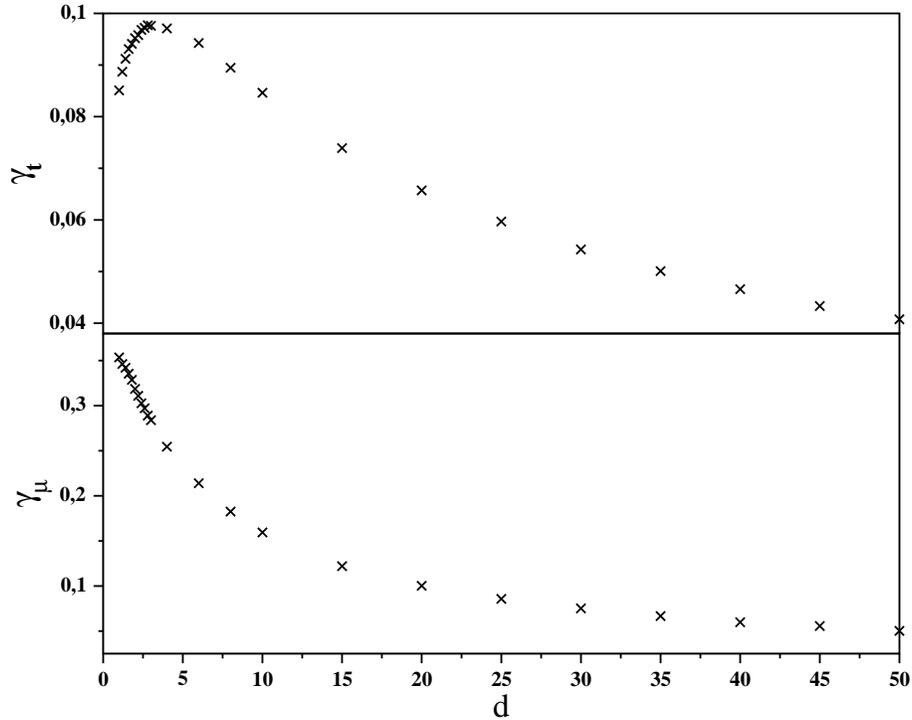}}
\caption{Escape rate $\gamma_\mu\,\,(\,\gamma_t\,)$ from two identical black holes ($M_1=M_2=1$) for different separation 
distance between them. The black holes are located at $z_{bh}=\pm d$. The data were obtained using ensembles of photons with 
initial velocity parallel to the x-axis and initial positions uniformly distributed along the $z$-axis.} 
\label{nt3}
\end{figure}

The escape rate of photons depends on the masses of the black holes and the separation distance between them.
The larger are the masses the more extensive in space are the unstable periodic orbits that form part of the fractal repeller 
that marks the boundary between dynamical stability and instability, and therefore the regions where the light rays fall into 
the black holes. It could be expected then that the escape rate of photons decreases as the masses of the black holes become 
larger. Figure (\ref{nt2}) illustrates this behavior and shows the decrease of the escape rate as the mass of one of the 
black holes is increased. Figure (\ref{nt3}) shows the dependence of the escape rate on the separation distance between 
the two black holes. In both cases, after an initial transient, the escape rate decays asymptotically with the distance $d$.
A similar behavior is observed in other scattering systems where the escape rate decays as $f(d)/d$, being $f(d)$ a function
determined by the dispersion of the scattering interaction potential and $1/d$ is the contribution of the time of flight 
\cite{gaspard1992,gaspard1998}.


In figure (\ref{nt3}) we consider the scattering of photons from two black holes with masses  $M_1=M_2=M$, located at 
different positions ($0,\pm d$). From (\ref{funcU}) and (\ref{eqnmotion},\ref{eqnmotion2}) it can be easily shown that 
a system of two black holes with scaled masses ${\tilde m}_1={\tilde m}_2=M/d$ located at $(0,\pm 1)$ satisfies identical 
dynamics with respect to an scaled time $\tilde\tau$ defined by $d\alpha/d\tau=d{\tilde\alpha}/d\tilde\tau\,
(d\tilde\tau=d\tau/d)$, being ${\tilde\alpha}=\alpha/d\,(\alpha=x,z)$ scaled coordinates. Hence if we denote $\gamma_d$ the 
escape rate from two identical black holes of mass $M$ located at ($0,\pm d$), and ${\tilde \gamma}$ the escape rate from 
two black holes of the same mass  ${\tilde m}=M/d$ located at $(0,\pm 1)$, it follows that $\gamma_d={\tilde \gamma}/d$.

\section{Escape dynamics of photons from a perturbed black hole}

In this section we study the escape rate from an extreme Reissner-Nordstr\"om black hole of mass $M_1$ that is
slightly perturbed by the interaction with an extreme Reissner-Nordstr\"om black hole of mass $M_2$ located at distance $d$.
In polar coordinates, the Hamiltonian can be written as
\beq\label{pert1}
{\cal H}=-\frac{h}{U^2} = -\frac{1}{U^2}\left[
\frac{1}{2}\left(p_r^2+\frac{p_\theta^2}{r^2}\right)-\frac{U^4}{2}\right]
\eeq
with the function $U$ given by:
\beq\label{pert2}
U=1+\frac{M_1}{r}+\frac{M_2}{\sqrt{r^2+d^2-2rd\sin\theta}}
\eeq

The analysis of the dynamics of this system can be greatly simplified if we consider the evolution with respect to a {\sl time} 
$\nu$ defined by the transformation
\begin{equation}
\frac{d\,}{d\mu}=-\frac{1}{U^2}\frac{d\,}{d\nu}.
\end{equation}
Thus we focus on the equations of motion associated with the Hamiltonian $h$,
\beq\label{pert4}
\frac{dX}{d\nu}=\Sigma\cdot\frac{\partial h}{\partial X}
\hspace*{1.3cm}X=(r,\theta,p_r,p_\theta)
\eeq
In these equations, $h$ can be analyzed as a kinetic energy term plus an interaction term $-U^4/2$. 
The function $U^4$ may be expressed as an expansion in powers of $(d/r)$, with the first two terms given by:  
\beq\label{pert5}
U^4=\left(1+\frac{M_1+M_2}{r}\right)^4+\frac{4M_2d}{r^2}\left(1+\frac{M_1+M_2}{r}\right)^3\sin\theta + {\cal O}
\left[\left(\frac{d}{r}\right)^2\right]
\eeq

In the limit $(M_1+M_2)^2>>2M_2d$ the dynamics of the system is dominated by the radial term in the expansion 
(\ref{pert5}), which is perturbed by the remaining angular terms. Thus to zero order approximation the space time associated 
with the two extreme black holes becomes spherically symmetric. There exists a single unstable periodic orbit which is a 
circle of radius $r=M\equiv(M_1+M_2)$ and period $T_\nu=M\pi/2$. The time evolution along this orbit is given by
\beqa\label{cir_orb1}
r_s(\nu)=M \hspace*{0.6cm}&&\hspace*{0.6cm} \theta_s (\nu)=\theta_0+\frac{4\nu}{M}   \\  \nonumber
p_{rs}(\nu)=0 \hspace*{0.6cm}&&\hspace*{0.6cm}  p_{\theta s}(\nu) = L
\eeqa
with angular momentum $L=4M$. The periodic orbits associated with the affine parameter and the killing time 
are also a circle of radius $r=M$, but with periods $T_\mu=2\pi M$ and $T_t=8\pi M$ respectively.

To zero order approximation the escape rate of photons from an {\sl effective} black hole of mass $M$ can be determined 
from the analysis of the linear stability of the circular orbit (\ref{cir_orb1}), 
$X_s\equiv (r_s, \theta_s, p_{rs},  p_{\theta s})$. This implies the study of the time evolution of a small perturbation, 
$\delta X\equiv(\delta r, \delta\theta,\delta p_r,\delta p_\theta)$, with respect $X_s$. In terms of the 
Hamiltonian $h$, the linear differential equations that describe this evolution, $d\,(\delta X)/d\nu={\sf l}(\nu)\cdot\delta X$  
with ${\sf l}(\nu)= \Sigma \cdot \frac{\partial^2 h}{\partial X^2}|_{X_s}$, are:
\beqa\label{cir_orb2}
\frac{d\,}{d\nu}
\left(\begin{array}{c}
\delta {r} \\ \delta {\theta} \\ \delta {p_r} \\ \delta {p_\theta}
\end{array}\right)=
\left(\begin{array}{cccc}
0 & 0 & 1 & 0 \\
-\frac{8}{M^2} & 0 & 0 & \frac{1}{M^2} \\
\frac{8}{M^2} & 0 & 0 & \frac{8}{M^2} \\
0 & 0 & 0 & 0 \end{array} \right) \cdot
\left(\begin{array}{c}
 \delta r \\ \delta \theta \\  \delta p_r \\ \delta p_\theta
\end{array}\right)
\eeqa
and the fundamental matrix ${\sf M}(X_s,\nu)$, solution of ${\sf M}'(X_s,\nu)={\sf l}(\nu)\cdot{\sf M}(X_s,\nu)$ with 
${\sf M}(X_s,0)={\sf 1}$, is given by 
\beqa\label{matrix_M}
{\sf M}(X_s,\nu)=\left(\begin{array}{cccc}
C(\nu) & 0 & \frac{M}{2\sqrt{2}}S(\nu) & \left[C(\nu)-1\right] \\
-\frac{2\sqrt{2}}{M}S(\nu) & 1 & \left[1-C(\nu)\right] & 
\left[\frac{9}{M^2}\nu-\frac{2\sqrt{2}}{M}S(\nu)\right] \\
\frac{2\sqrt{2}}{M}S(\nu) & 0 & C(\nu) & \frac{2\sqrt{2}}{M}S(\nu) \\
0 & 0 & 0 & 1 
\end{array}\right)
\eeqa
with
\beq
C(\nu)=\cosh\left(\frac{2\sqrt{2}}{M}\nu\right)
\hspace*{1cm}{\mbox {and}}\hspace*{1cm}
S(\nu)=\sinh\left(\frac{2\sqrt{2}}{M}\nu\right)
\eeq
The stretching factors associated with the circular orbit are obtained from the eigenvalues
of the matrix ${\sf M}(X_s,\nu)$, 
\beq
\left\{ 1 , 1 , e^{-\frac{2\sqrt{2}}{M}\nu} , e^{\frac{2\sqrt{2}}{M}\nu} \right\},
\eeq
evaluated at one primitive period $T_\nu$. That is,
\beq
\Lambda= \left\{ 1 , 1 , e^{-\sqrt{2}\pi} , e^{\sqrt{2}\pi} \right\}.
\eeq
We recall that these stretching factors are invariant under time transformations, and therefore are
identical for the affine parameter and the killing time.
As expected, two of them are equal to one. The values $e^{-\sqrt{2}\pi}$ and $e^{\sqrt{2}\pi}$ 
are related to the stable and unstable manifolds of the unique unstable periodic orbit respectively. 
The Lyapunov exponents associated with this orbit are given by 
\beq	
\lambda_\mu=\frac {\ln |\Lambda|}{T_\mu} = 
\left\{ 0 , 0 , -\frac{1}{\sqrt{2}M}, \frac{1}{\sqrt{2}M} \right\}
\eeq
in the time evolution dictated by the affine parameter, and 
\beq
\lambda_t=\frac {\ln |\Lambda|}{T_t} =
\left\{ 0 , 0 , -\frac{1}{4\sqrt{2}M}, \frac{1}{4\sqrt{2}M} \right\}
\eeq
in the evolution with respect to the killing time.
Thus to zero order the escape rate of photons from the unstable circular periodic orbit around the effective 
extreme black hole of mass $M$  is determined by the leading exponent
$\gamma_\mu^{(0)}=1/\sqrt{2}M$ and $\gamma_t^{(0)}=1/4\sqrt{2}M=\gamma_\mu^{(0)}/4$.

In this system there is a unique unstable periodic. Since the dynamics on the 
repeller is regular, the Kolmogorov-Sinai entropy is zero, $h_{KS}=\lambda-\gamma^{(0)}=0$. Thus the system 
is hyperbolic but nonchaotic. The Pollicott-Ruelle resonances $\sigma_{pr}$ are given by the zeroes of the Zeta 
function (\ref{zf}) associated with the single unstable periodic orbit,
\beq
Z(\sigma_{pr})=\prod_{k=0}^{\infty}\left(1-\frac{e^{-\sigma_{pr} T}}{|\Lambda|\Lambda^k}\right)^{k+1}=0,
\eeq
being $T$ the period $T_\mu$ or $T_t$. That is,
\beq\label{resonances_mu}
\sigma_{pr}^{(\mu)} (k,n)=-\frac{(k+1)}{\sqrt{2}M}+i\frac{n}{M}
\eeq
for the affine parameter and 
\beq\label{resonances_t}
\sigma_{pr}^{(t)} (k,n)=-\frac{(k+1)}{4\sqrt{2}M}+i\frac{n}{4M}
\eeq
for the killing time, with $k=0,1,2,...$ and $n=0,\pm 1, \pm 2,...$  
The resonances belong to the lower half plane of the complex plane $\sigma$ (Re $\sigma_{pr}<0$) and, 
as occurs in the two-disk scatterer \cite{gaspard1998,gaspard2002}, their spectrum forms a semi-infinity periodic 
array. Here the spacing along the (Re $\sigma$)-axis is given by the escape rate $\gamma_\mu^{(0)}=1/\sqrt{2}M$ 
$(\gamma_t^{(0)}=1/4\sqrt{2}M)$, and by the frequency $w_{\mu}=1/M$ $(w_{t}=1/4M)$ along the (Im $\sigma$)-axis.
These complex resonances play an important role in the time evolution of an statistical ensemble of photons; they 
determine the different decay modes and their frequencies in a typical scattering process. 
The ensemble dynamics is ruled by the resonances which are the closest to the imaginary axis $(k=0)$. The real
part of these leading resonances controls the exponential decay on the longest time scale, which defines the 
escape rate of the system; and their imaginary parts give the frequencies of the oscillations that appear
superimposed on the gross exponential decay. 

We now study the effect of the first angular term in the expansion (\ref{pert5}) on the zero-order circular orbit 
(\ref{cir_orb1}) associated with the leading radial term. We still assume the limit $M^2=(M_1+M_2)^2>>2M_2d$, 
in which this term can be treated as an perturbation to the spherically symmetric motion. 
To make a first order perturbative analysis we write the $U^4$ function in the interaction term in the form
\beq\label{pert6}
U^4=\left(1+\frac{M}{r}\right)^4+\varepsilon\frac{4M_2d}{r^2}\left(1+\frac{M}{r}\right)^3\sin\theta
\eeq
where we have introduced a perturbative parameter $\varepsilon$, which is set equal to one at the end of the 
analysis. The perturbed trajectory can be written as
\beqa\label{pert7}
& &r_1(\nu)=r_s(\nu)+\varepsilon r_c(\nu) \hspace*{0.6cm},\hspace*{0.6cm} 
p_{r1}(\nu)=p_{rs}(\nu)+\varepsilon p_{rc}(\nu)  \nonumber \\
& &\theta_1(\nu)=\theta_s(\nu)+\varepsilon \theta_c(\nu) \hspace*{0.6cm},\hspace*{0.6cm}
p_{\theta 1}(\nu)=p_{\theta s}(\nu)+\varepsilon p_{\theta c}(\nu)
\eeqa
with first-order corrections to the circular orbit (\ref{cir_orb1}) that satisfy the equations 
\beqa\label{pert8}
& & \frac{dr_c}{d\nu}=p_{rc}   \nonumber \\
& & \frac{d\theta_c}{d\nu}=\frac{1}{M^2}\left(p_{\theta c}-8r_c\right) \nonumber \\
& & \frac{dp_{rc}}{d\nu}=-\frac{56M_2d}{M^3}\cos\left(\frac{4\nu}{M}\right)+\frac{8}{M^2}\left(p_{\theta c}+r_c\right)
\nonumber \\
& & \frac{dp_{\theta c}}{d\nu}=-\frac{16M_2d}{M^2}\sin\left(\frac{4\nu}{M}\right).
\eeqa
The periodic solution of these equations gives the first-order perturbed orbit,
\beqa\label{pert9}
& & r_1(\nu)=M+\frac{dM_2}{M}\cos\left(\frac{4\nu}{M}\right) \nonumber \\
& & \theta_1(\nu)=\theta_0+\frac{4}{M}\nu-\frac{dM_2}{M^2}\sin\left(\frac{4\nu}{M}\right) \nonumber \\
& & p_{r1}(\nu)=-\frac{4dM_2}{M^2}\sin\left(\frac{4\nu}{M}\right) \nonumber \\
& & p_{\theta 1}(\nu)=4M+\frac{4dM_2}{M}\cos\left(\frac{4\nu}{M}\right).
\eeqa

\begin{figure}[h]
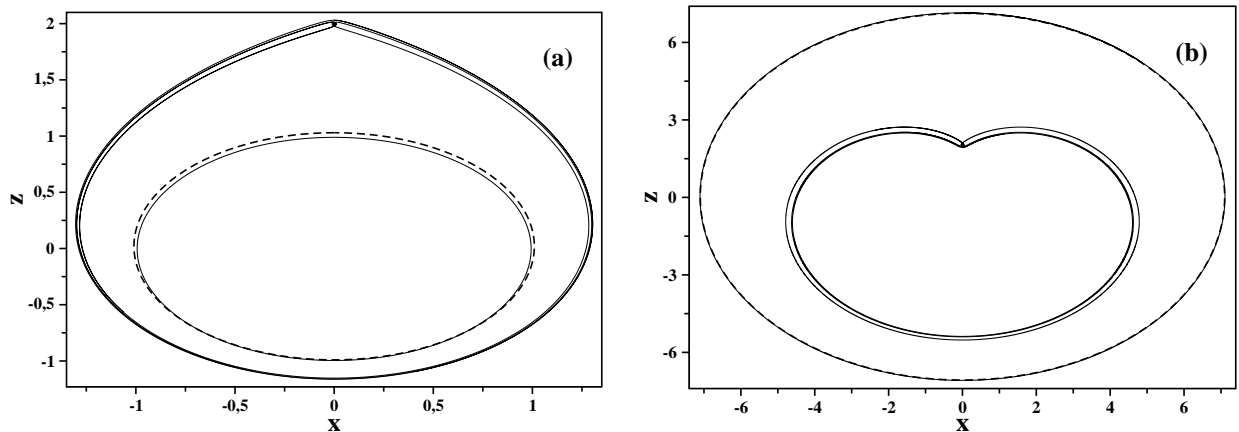

\includegraphics[width=0.48\textwidth]{case_A.eps}
\hspace*{0.3cm}
\includegraphics[width=0.48\textwidth]{case_D.eps}
\caption{Some of the periodic orbits in two systems of a black hole of mass $M_1$ perturbed by the interaction with 
a black hole of mass $M_2$ located at distance $d$. The solid lines give the periodic orbits $\{++\}$, $\{--\}$, $\{+-\}$, 
$\{+-++\}$ and $\{+\circ -\circ\}$, and the dashed line is the first-order periodic orbit (\ref{pert9}). The system 
$(a)$ corresponds to $(\,M_1=1,\,M_2=0.01,\,d=2\,)$ and $(b)$ to $(\,M_1=7,\,M_2=0.1,\,d=2)$. In both systems the 
main black hole of mass $M_1$ is located at $(0,0)$ and the black hole of mass $M_2$ at $(0,d)$.}
\label{pert_orb}
\end{figure}

\begin{figure}[h]
\includegraphics[width=0.75\textwidth]{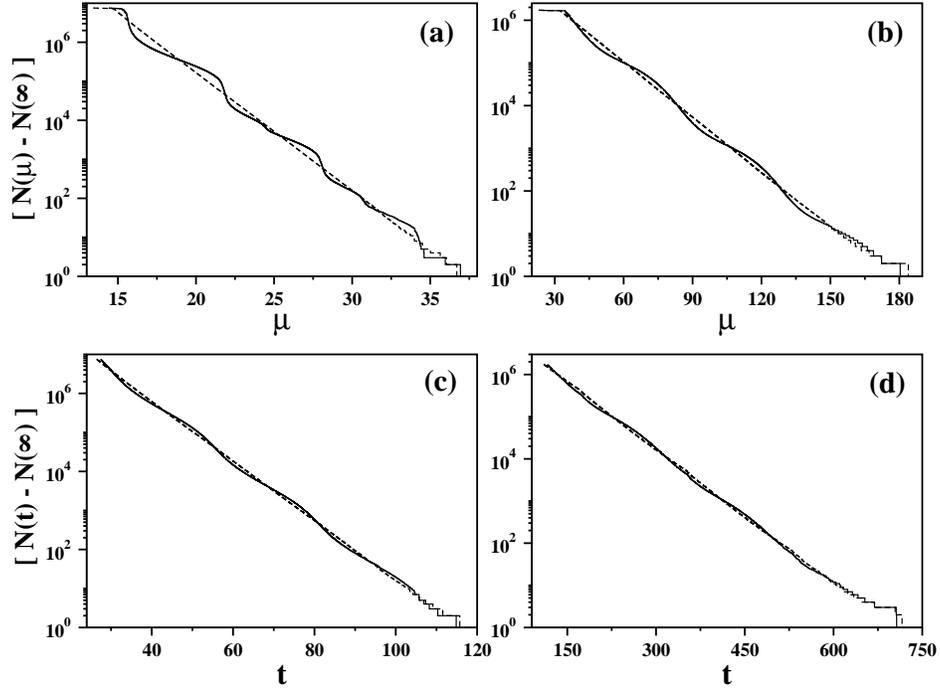}
\vspace*{0.1cm}
\caption{The evolution of $[N(\mu)-N(\infty)]\,\,([N(t)-N(\infty)])$ with the affine parameter (killing time) for two 
statistical ensembles in the perturbed systems of figure (\ref{pert_orb}). Panels $(a)$ and $(c)$ correspond to the system
$(a)$ in figure (\ref{pert_orb}), and panels $(b)$ and $(d)$ to the system $(b)$ in figure (\ref{pert_orb}).
The solid line gives the escape from a circle of control centered in between the two black holes, at $(0,d/2)$, and the dashed 
lines the escape from a circle of control centered on the main black hole, at $(0,0)$. See the Section VII for comments.}
\label{pert_osc}
\end{figure}

This is a quasi-circular orbit that nearly reproduces the periodic orbit $\{+-\}$ in two black-hole 
systems in which $(M_1+M_2)>d$, and the periodic orbit $\{--\}\,(\{++\})$ in systems where $(M_1+M_2)<d$, 
see figure (\ref{pert_orb}). 
The more complex periodic orbits (longer symbolic coding) are out of scope of our perturbative analysis. However, 
in the limit $(M_1+M_2)^2>>2M_2d$  these are highly unstable periodic orbits that play a secondary role in the escape 
rate of photons from the black holes. Here the escape rate is controlled by the quasi-circular periodic orbit, 
$\{+-\}$ or $\{--\}\,(\{++\})$, which results from the slight deformation of the circular orbit of radius $M$.
Hence the analysis of the linear stability of the zero- and first-order periodic orbits, (\ref{cir_orb1}) and
(\ref{pert9}), should provide a good estimate of the escape rate of photons from a perturbed black hole. Indeed, as the 
results in Table (\ref{tab_pert}) indicate, both the analytical escape rate $\gamma_\mu^{(0)}\,\,(\,\gamma_t^{(0)}\,)$ derived 
from the analysis of the linear stability of the zero-order circular orbit (\ref{cir_orb1}) and the escape rate 
$\gamma_\mu^{(1)}\,\,(\,\gamma_t^{(1)}\,)$  obtained from the leading eigenvalue of the fundamental matrix associated with 
the first-order quasi-circular orbit (\ref{pert9}) are very close to the escape rate $\gamma_\mu^{(num)}\,\,(\,\gamma_t^{(num)}\,)$ 
calculated from the numerical simulations that consider an statistical ensemble of photons, see figure (\ref{pert_osc}).  
\begin{table}
\begin{tabular}{|c||c|c|c||c|c|c|} \hline
 System &  $\gamma_\mu^{(0)}$ &  $\gamma_\mu^{(1)}$ & $\gamma_\mu^{(num)}$ & 
 $\gamma_t^{(0)}$ &  $\gamma_t^{(1)}$ & $\gamma_t^{(num)}$\\ \hline\hline
$\,\,(\,M_1=1,\,M_2=0.01,\,d=2\,)\,\,$ & $\,\,0.70011\,\,$ &  $\,\,0.69559\,\,$ & $\,\,0.69258\,\,$ 
& $\,\,0.17503\,\,$ &  $\,\,0.17390\,\,$ & $\,\,0.17624\,\,$ \\ \hline
$\,\,(\,M_1=7,\,M_2=0.1,\,d=2\,)\,\,$ & $\,\,0.09959\,\,$ &  $\,\,0.09957\,\,$ & $\,\,0.09901\,\,$  
& $\,\,0.02490\,\,$ &  $\,\,0.02489\,\,$ & $\,\,0.02436\,\,$ \\ \hline
\end{tabular}
\caption{Escape rate values for two systems of perturbed black holes. $\gamma_\mu^{(0)}\,\,(\,\gamma_t^{(0)}\,)$ is the 
escape rate derived from the zero-order circular orbit (\ref{cir_orb1}), $\gamma_\mu^{(1)}\,\,(\,\gamma_t^{(1)}\,)$ is the 
escape rate obtained from the analysis of linear stability of the first-order orbit (\ref{pert9}) and 
$\gamma_\mu^{(num)}\,\,(\,\gamma_t^{(num)}\,)$ is the escape rate calculated from the numerical simulations that 
consider an statistical ensemble of photons.}
\label{tab_pert}
\end{table}
As figure (\ref{nt2}) shows, the zero-order analytic estimate $\gamma_\mu^{(0)}\,\,(\,\gamma_t^{(0)}\,)$ even provides an 
approximated value for the escape rate from a system of two black holes with similar masses. 

\begin{figure}[h]
\includegraphics[width=0.75\textwidth]{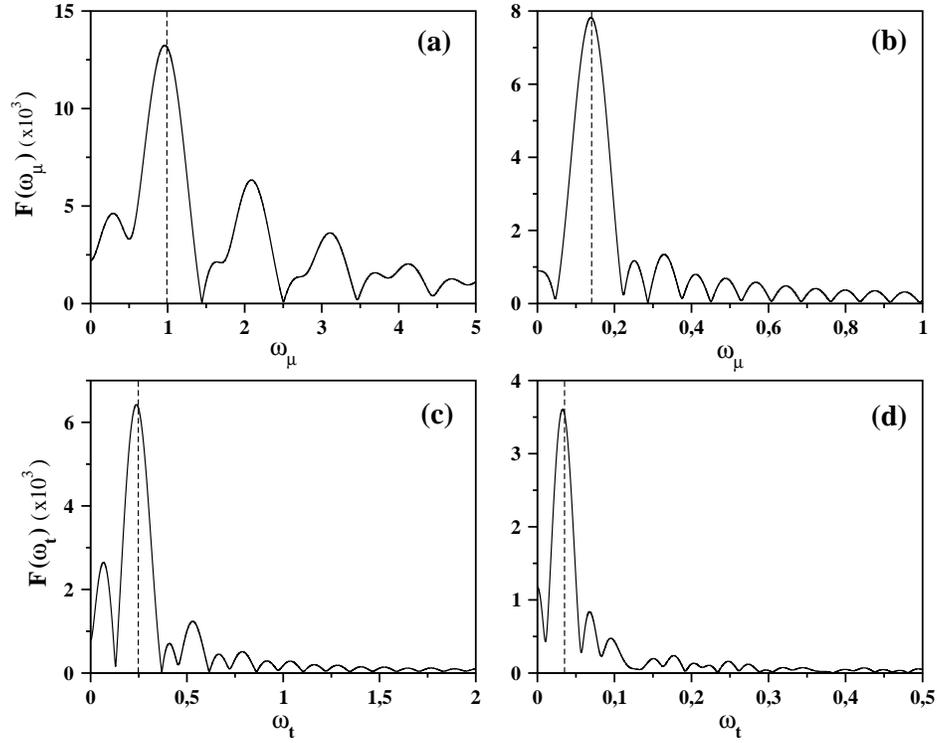}
\vspace*{0.1cm}
\caption{The Fourier transform of the four signals in figure (\ref{pert_osc}). The dashed lines localize the frequency of 
the main oscillation in the escape rate predicted by the leading zero-order Pollicott-Ruelle resonances, $w_\mu\simeq 1/M\,\,
(\,w_t\simeq 1/4M\,)$.}
\label{fourier}
\end{figure}

In the numerical simulations with an statistical ensemble of photons, see figure (\ref{pert_osc}), the escape from the perturbed
black hole presents a smooth oscillation superposed on the gross exponential decay determined by the escape rate 
$\gamma_\mu^{(num)}\,\,(\,\gamma_t^{(num)}\,)$. 
The zero-order Pollicott-Ruelle resonances (\ref{resonances_mu}) and (\ref{resonances_t}) account for this behavior of the decay 
curves; the exponential decay determined by the escape rate $\gamma_{\mu}\simeq \gamma_\mu^{(0)}\,\,(\,\gamma_t\simeq \gamma_t^{(0)}\,)$ 
is given by the real part of the leading resonances $k=0$, and the frequency of the oscillation 
$w_\mu\simeq 1/M\,\,(\,w_t\simeq 1/4M\,)$ is given by the imaginary part of the leading resonance given by $n=1\,(k=0)$. 
This is illustrated in figure (\ref{fourier}) which displays the Fourier spectra of the signals in figure (\ref{pert_osc}). It
clearly shows that the most intense peaks in the spectra, associated with the frequency of the main oscillation in the escape rate, 
are localized pretty close to these frequencies predicted by the leading zero-order Pollicott-Ruelle resonances. These peaks 
correspond to the frequency of the leading orbits, the orbit $\{--\}(\{++\})$ in case A and the orbit $\{+-\}$ in case B, see figure 
(\ref{pert_orb}).
This is another indication of the leading role of the quasi-circular orbit that results from the slight deformation of the 
circular orbit of radius $r=M$ in the escape dynamics of photons.

\section{Conclusions}

We have studied the scattering of photons in the Majumdar-Papapetrou static spacetime of two extreme Reissner-Nordstr\"om black 
holes held fixed in space due to the balance between the gravitational attraction of their masses and the electrostatic repulsion
of their charges. We have identified the set of unstable periodic orbits that form part of the fractal repeller which fully 
characterizes the chaotic escape dynamics of the photons from the two black holes. These orbits were classified according to a 
symbolic coding. 

The linear stability of the dynamics, in particular the unstable periodic orbits, was analyzed using their stretching factors, 
which are given by the eigenvalues of the fundamental matrix integrated up to one primitive period of the orbits. With the 
primitive periods and stretching factors of the periodic orbits we determined the topological pressure, and from this function and 
its first derivative the main quantities of chaos that characterize the fractal repeller. In particular the escape rate, which was 
calculated using the trace formula for hyperbolic flows of Cvitanovic and Eckhardt. This escape rate derived from the periodic 
orbit theory was in good agreement with the value obtained from numerical simulations that analyze the escape dynamics of 
an statistical ensemble of photons.

From the study of the linear stability of the dynamics we also identified the periodic orbit that plays the leading role
in the escape of photons from the two black-hole configuration. In systems with two identical extreme Reissner-Nordstr\"om black 
holes the escape is mainly controlled by the outermost periodic orbit $\{+-\}$ that encircles the two black holes.
In systems where a main black hole of mass $M_1$ is perturbed by the weak interaction with a black hole of mass $M_2$, 
the escape of photons is dominated by the quasi-circular orbit that results from the slight deformation of the circular orbit
of radius $r=M=(M_1+M_2)$.
Depending on the separation distance between the two black holes this leading orbit can cover the whole system, and corresponds 
to the periodic orbit $\{+-\}$, or only encircles the main black hole, and is given by the periodic orbit $\{++\}\,(\{--\})$.

In systems of a perturbed black hole, the analysis of the linear stability of the unique zero-order periodic orbit in a 
perturbative treatment provides an analytic estimate of the escape rate, 
$\gamma_\mu^{(0)}=1/\sqrt{2}M\,\,(\,\gamma_t^{(0)}=1/4\sqrt{2}M\,)$, which is in good agreement with numerical value derived 
from the analysis of the linear stability of the first-order perturbative periodic orbit, and also with the escape rate obtained 
from the time evolution of an statistical ensemble of photons. 
This analytical value even provides an approximated value for the escape rate of photons from a system of two extreme 
Reissner-Nordstr\"om black holes with similar masses. And therefore a way to estimate the mass of a black hole from the escape
rate of photons.

\begin{acknowledgments}

D. Alonso thanks C.P. Dettmann for fruitful discussions. We thank G. Contopoulos and N.J. Cornish for pointing out important 
references. Financial support has been provided by Ministerio de Educaci\'on y Ciencia (FIS2004-05687, FIS2005-02886, 
FIS2007-64018) and Gobierno de Canarias (PI2004/025).

\end{acknowledgments}

\end{document}